\documentclass[12pt]{article}
\pdfoutput=1
\usepackage{amsthm,amsmath,latexsym,amssymb,amsfonts,amscd}
\usepackage{graphics,lscape,fancyhdr,array,stmaryrd,euscript}
\usepackage{ifthen,empheq,slashed}
\usepackage{color}
\numberwithin{equation}{section}
\usepackage{setspace}
\usepackage{cite}
\usepackage{mathrsfs}
\usepackage[hang,flushmargin]{footmisc}
\usepackage[colorlinks,
			breaklinks,
			hyperfootnotes,
			linktoc=page,
            linkcolor=blue,
            citecolor=blue,
            urlcolor=blue,
            unicode,
            pdfstartview=FitH,
            bookmarks,
            backref=page,
            bookmarksnumbered]{hyperref}
\renewcommand*{\backref}[1]{}
\renewcommand*{\backrefalt}[4]{[{\footnotesize
    \ifcase #1 Not cited.%
          \or Cited on page~#2.%
          \else Cited on pages #2.%
    \fi%
    }]}
\usepackage{makecell}
\usepackage{geometry}

\newgeometry{left=0.8in,right=0.8in,top=.8in,bottom=1.1in}
\usepackage[most]{tcolorbox}
\usepackage{enumitem}
\usepackage{footnotebackref}
\usepackage{stackengine}
\stackMath
\usepackage[retainorgcmds]{IEEEtrantools}
\usepackage{microtype}

\def\be {\begin{equation}}
	\def\ee {\end{equation}}
\def\bea {\begin{eqnarray}}
	\def\eea {\end{eqnarray}}
\def\bc {\begin{center}}
	\def\ec {\end{center}}
\def\bg {\begin{align}}
	\def\eg {\end{align}}
\def\bi {\begin{itemize}}
	\def\ei {\end{itemize}}

\def\le {\left}
\def\ri {\right}
\def\p {\partial}

\def\c  {\cdot}
\def\a  {\alpha}
\def\b  {\beta}
\def\g  {\gamma}
\def\d  {\delta}
\def\e  {\eta}

\def\m  {\mu}
\def\n  {\nu}
\def\s {\sigma}

\def\ep{\varepsilon}

\def\1{_{_1}}
\def\2{_{_2}}

\def\beak{\begin{IEEEeqnarray*}}
\def\eeak{\end{IEEEeqnarray*}}
\def\nk{\IEEEyesnumber\phantomsection}
\def\snk{\IEEEyessubnumber}

\newcommand\ad {{\dot{\alpha}}}
\newcommand\bd {{\dot{\beta}}}

\newcommand\thd {{\bar{\theta}}}
\renewcommand\th {{\theta}}
\newcommand\D {{\rm D}}
\newcommand\Dd {{\bar{\rm D}}}
\newcommand\pa {{\partial}}

\begin{document}

	\hfill
	\begin{flushright}
		{IPM/P-2023/61}
	\end{flushright}

 $\vspace{.5cm}$
	\begin{center}

		{ \resizebox {\linewidth} {\height} {\bfseries \LARGE{Super-Carrollian and Super-Galilean Field Theories}}} \\

		\vskip 0.05\textheight

  Konstantinos \textsc{Koutrolikos}\,\,$^{\color{purple}a}$, Mojtaba \textsc{Najafizadeh}\,\,$^{{\color{purple}b},\,{\color{purple}c}}$

		\vskip 0.01\textheight
{\raggedright

  \vspace*{15pt}
${}^{\color{purple}a}$~{ Department of Physics, University of Maryland, \\
     \,~~~College Park, Maryland 20742-4111, USA}

\vspace*{15pt}
${}^{\color{purple}b}$~{ School of Physics, Institute for Research in Fundamental Sciences (IPM), \\
     ~~~P.O.Box 19395-5531, Tehran, Iran }

\vspace*{15pt}
${}^{\color{purple}c}$~{ Department of Physics, Faculty of Science, Ferdowsi University of Mashhad, \\
     ~~~P.O.Box 1436, Mashhad, Iran } \\
     }

		\vskip 0.01\textheight

  \href{mailto:koutrol@umd.edu}{koutrol@umd.edu},
  \href{mailto:mnajafizadeh@ipm.ir}{mnajafizadeh@ipm.ir}

		\vskip 0.05\textheight

		{\bf Abstract }

	\end{center}
	\begin{quotation}
The exploration of scalar field theories that exhibit Carroll and Galilei symmetries has attracted a
lot of attention. In this paper, we generalize these studies to fermionic field theories and
construct consistent electric and magnetic descriptions of Carrollian and Galilean spin
$\tfrac{1}{2}$ fermions. We showcase various methods that offer complementary perspectives into the limiting process of the underlying relativistic theories. Moreover, we extend our study to
$\mathcal{N}=1$ off-shell supersymmetric field theories in four dimensions. By introducing suitable
Grassmann-analyticity conditions, we formulate the corresponding super-Carrollian and super-Galilean
theories. These theories combine the established Carroll/Galilei scalars with the Carroll/Galilei
fermions and a range of auxiliary fields into supermultiplets.
	\end{quotation}


	\newpage
	\tableofcontents
	\newpage
\section{Introduction}
The concept of nonrelativistic geometries\footnote{Geometries of smooth manifold where the local
Lorentz group has been replaced by some other nonrelativistic group.} is not a new one, as it was
introduced around a century ago by Cartan in his development of Newton-Cartan gravity
\cite{Cartan:1923zea} based on a torsionless non-Lorentzian geometry. Nevertheless, recently there
has been a significant interest in the exploration and study of non-Lorentzian theories driven by a
multitude of applications in a variety of physical systems (for reviews see
\cite{Henneaux:2021yzg,Oling:2022fft,Bergshoeff:2022eog,Bagchi:2022owq}).
These theories abide by a principle of relativity characterized by symmetries that emerge through the
breaking of Lorentz invariance. This breaking, leads to many possible kinematic algebras\footnote{For
a detailed analysis of classifications see
\cite{Bacry:1968zf,Bacry:1986pm,Figueroa-OFarrill:2017ycu,Figueroa-OFarrill:2018ilb}.},
but the main two structures are the (i) Carrollian and (ii) Galilean algebras. Both of these algebras
are well-known contractions of the relativistic Poincar\'{e} algebra
\cite{Inonu:1953sp,Lévy1965,1966NCimA..44..512S} associated with small or large characteristic
velocities compared with the speed of light or equivalently with the $c\to 0$ or $c \to \infty$
limits respectively.

\vspace{.3cm}

There are several motivations that encourage the study of these non-Lorentzian theories and their application to various physical systems. Nevertheless, our inspiration stems from developments in (a) flat space holography and (b) nonrelativistic/Carrollian string theory. It is a fact that the Carrollian conformal algebra is isomorphic to the BMS algebra in one dimension higher \cite{Duval:2014uva} and therefore relevant for the celestial approach to flat space holography \cite{Barnich:2009se,Bagchi:2010zz,Bagchi:2012cy,Donnay:2022aba,Donnay:2022wvx,Saha:2023hsl,Saha:2023abr}. Moreover, Carrollian string theories \cite{Cardona:2016ytk} as well as nonrelativistic string theories\cite{Gomis:2000bd,Danielsson:2000gi,Gomis:2005pg} and tensionless string theories \cite{Isberg:1993av,Bagchi:2015nca,Bagchi:2020fpr,Bagchi:2016yyf,Bagchi:2017cte,Bagchi:2018wsn} are known examples of solvable string theories and provide a unitary and ultraviolet complete framework that connects different corners of string theory. Naturally, this creates a resurgence of interest in non-Lorentzian geometries and quantum field theories that arise from such string theories.

\vspace{.3cm}

At the heart of defining nonrelativistic superstring theory lies nonrelativistic supersymmetry.
The non-Lorentzian supersymmetry that emerges within this context is considerably less developed (see early works
\cite{Puzalowski:1978rv,deAzcarraga:1991fa,Bergman:1995zr,Bergman:1996bx}) and
not much progress has been made in the direction of
nonrelativistic superspace and nonrelativistic supermultiplets.
The purpose of this paper is to construct Carrollian and Galilean supersymmetric theories of matter. The bosonic parts of
these theories have been explored in detail in \cite{Bergshoeff:2022eog}. It has been shown that in addition to the
naturally defined (1) electric (time derivative dominated) Carroll-invariant scalar theory and (2) magnetic (space
derivative dominated) Galilei-invariant scalar theory, one can construct by application of the ``\emph{seed}'' Lagrangian
method alternative magnetic Carroll-invariant and electric Galilei-invariant scalar theories. This method is based on the
fact that Carrollian and Galilean boosts are nilpotent operators. These alternative descriptions can be also understood
as the appropriate (Carrollian or Galilean) limits of a dual relativistic theory \cite{Gomis:2005pg}, where the divergent
quadratic terms have been tamed by the introduction of a Lagrange multiplier.

\vspace{.3cm}

In this paper we introduce the spin 1/2 fermionic counterparts to the previously mentioned scalar theories \cite{Henneaux:2021yzg,deBoer:2021jej,Bergshoeff:2022qkx}. Through a range of methodologies, including Hamiltonian formulation, field expansion, and seed Lagrangian, we develop Carroll and Galilei fermionic field theories that
exhibit invariance under their respective boost transformations. Similar to the scalar theories, each fermionic theory has two formulations - one electric and one magnetic. Carrollian fermions have been introduced in \cite{Banerjee:2022ocj,Bagchi:2022eui}, however in these works the authors absorb the In\"{o}n\"{u}-Wigner contraction parameter in the background spacetime metric.
Consequently, at the appropriate contraction limit there is a significant alteration in the structure of the Clifford algebra and the structure of spinors. In contrast, we absorb the contraction parameter by appropriate coordinate or field rescalings that do not impact the $\gamma$ matrices. This approach proves advantageous as it enables the exploration of the supersymmetric extensions of these Carroll and Galilei theories.

\vspace{.3cm}

This is the focus of the second part of the paper where we embed the known Carroll/Galilei scalars together with their corresponding fermions within non-Lorentzian supermultiplets. There are two types of supersymmetries that can be imposed, contingent on whether the supersymmetry algebra closes under temporal translations (\emph{C-supersymmetry}) or spatial translations (\emph{G-supersymmetry}). For each type of supersymmetry, we develop the appropriate superspace description and use it in order to construct manifestly supersymmetric theories of matter that remain invariant under Carrollian or Galilean boosts. By introducing the notion of a \emph{C-chiral} superfield, we are able to describe in superspace a \emph{C-Wess-Zumino} model. This model provides the C-supersymmetric extension of electric Carroll matter, meaning that the space and time component description of the theory includes two electric Carroll scalars and an electric Carroll fermion. Similarly, we introduce a \emph{G-chiral} superfield and use it to construct the \emph{G-Wess-Zumino} model, which serves as the G-supersymmetric extension of magnetic Galilei matter which includes two magnetic Galilei scalars and one magnetic Galilei fermion. Furthermore, employing the seed Lagrangian approach, we utilize the G-Wess-Zumino and C-Wess-Zumino models as seeds to derive (1) a magnetic Carroll  theory with G-supersymmetry and (2) an electric Galilei theory with C-supersymmetry.

\vspace{.3cm}

The paper is structured as follows: In Section \ref{super carroll algebra}, we delve into the contractions of the $4D,~\mathcal{N}=1$ super-Poincar\'{e} algebra leading to the super-Carroll and super-Galilei algebras. Moreover, using the group manifold approach we (\emph{a.}) define suitable bosonic and fermionic superspace coordinates, (\emph{b.}) derive their transformations under supersymmetry and boosts and (\emph{c.}) define the differential operators that realize the action of supersymmetry and boost generators on the space of (super)fields as well as the necessary superspace spinorial derivatives. In Section \ref{Cfermions}, we develop electric and magnetic spin 1/2 fermionic theories that are invariant under Carroll boosts. Moving on to Section \ref{Gfermions}, we repeat the process for fermionic theories invariant under Galilei boosts. Sections \ref{sec5} and \ref{sec6} discuss the supersymmetric extension of electric and magnetic Carroll theories as well as electric and magnetic Galilei theories. In Appendices \ref{Carroll scalar rev} and \ref{Gscalar}, we offer a comprehensive review of all established methodologies used to construct the known electric and magnetic descriptions of Carroll and Galilei scalars. Also, for completeness, we discuss in Appendix \ref{appD} the formal off-shell supersymmetric extension of the ``\emph{simplest}'' theory for magnetic Carroll fields as defined in \cite{deBoer:2023fnj}.

\section{Super-Carroll and Super-Galilei Algebras} \label{super carroll algebra}
We start with the $4D, \mathcal{N}=1$ super-Poincar\'{e} algebra generators $\mathcal{J}_{mn}, \mathcal{P}_m,
\mathcal{Q}_{\a}, \mathcal{\bar{Q}}_{\dot{\alpha}}$ and introduce a contraction parameter $c$ by defining new generators
\beak{lll}
K_i \coloneqq c^{s_0} \mathcal{J}_{0i}~,~\quad & P_0 \coloneqq c^{r_0} \mathcal{P}_0~,~\quad & Q_{\a} \coloneqq c^{k} \mathcal{Q}_{\a}~,\\[5pt]
J_{ij} \coloneqq c^{s_1} \mathcal{J}_{ij}~,~\quad & P_{i} \coloneqq c^{r_1} \mathcal{P}_{i}~,~\quad & \bar{Q}_{\dot\alpha} \coloneqq c^k
\mathcal{\bar{Q}}_{\dot\alpha}~,\nk
\eeak
for some exponents $s_0, r_0, s_1, r_1, k$. Notice that the exponents associated with time components
are in general different than those associated with spatial components and thus allow for Lorentz symmetry
breaking. Also the exponents of $Q_{\a}$ and $\bar{Q}_{\dot\alpha}$ are the same due to Hermiticity.
The nonvanishing parts of the algebra of these generators are:
\beak{lll}
[K_i, K_j] = -\,i c^{2s_0-s_1}~ J_{ij}~,~~~~&[K_i,J_{jk}] = -\,ic^{s_1}~\d_{i[j}~K_{k]}~,~~~~&
[J_{ij},J_{kl}] = ic^{s_1}(\d_{[i|k|}J_{j]l} - \d_{[i|l|}J_{j]k})~,\\[8pt]
{}[K_i,P_0] = -\,ic^{s_0+r_0-r_1}~P_i~,~~~~&[K_i,P_j] = -\,ic^{s_0+r_1-r_0}~\d_{ij}P_0~,~~~~&
[J_{ij},P_k] = ic^{s_1}~\d_{[i|k|}P_{j]}~,\\[8pt]
{}[K_i,Q_\a] = c^{s_0}~(\s_{0i})_{\a}{}^{\b}~Q_{\b}~,~~~~&
[J_{ij},Q_\a] = c^{s_1}~(\s_{ij})_{\a}{}^{\b}~Q_\b~,~~~~&
[K_i,\bar{Q}_{\dot\a}] = c^{s_0}~(\bar{\s}_{0i})_{\dot\a}{}^{\dot\b}~\bar{Q}_{\dot\b}~,\\[8pt]
{}[J_{ij},\bar{Q}_{\dot\a}] = c^{s_1}~(\bar{\s}_{ij})_{\dot\a}{}^{\dot\b}~\bar{Q}_{\dot\b}~,~~~~&
\{Q_\a,\bar{Q}_{\dot\a}\} = -\,c^{2k-r_0}~(\s^0)_{\a\dot\a}~P_{0}&-\,c^{2k-r_1}~(\s^i)_{\a\dot\a}~P_i~.\nk
\eeak
In general, depending on the values of the various exponents there can be three types of terms: (i) $c$
independent terms, (ii) $c^n,~n>0$ terms and (iii) $\frac{1} {c^n},~n>0$ terms.
In order for this algebra to be well defined when we take the limit of the contraction parameter to zero
(Carroll) all terms must be of type (i) or type (ii). Similarly, when we take the limit of the contraction
parameter to infinity (Galilei) all terms must be of type (i) or type (iii). Moreover, we require
the
survival of the algebra of spatial rotations ($s_1=0$) and we explore algebras that deviate
from the starting algebra ($s_0\neq0$). With all that in mind, we find two
distinct, nontrivial
consistent algebras that correspond to the following choices:
\beak{l}\label{CG}\nk
\text{Carroll:}~~s_0+r_0-r_1\neq0~,~\quad s_0+r_1-r_0=0~,~\quad 2k=r_0~,~\snk\\[8pt]
\text{Galilei:}~~\,s_0+r_0-r_1=0~,~\,\quad s_0+r_1-r_0\neq0~,~\quad 2k=r_1~.~\snk
\eeak
In order for the generators of these algebras to behave properly in the appropriate $c$ limit
\beak{l}\nk
P_0 ~\propto ~c^{r_0}~\frac{1}{c}\,\frac{\partial}{\partial t} =c^{r_0-1}~\frac{\partial}{\partial t}~,\snk\\[5pt]
P_i ~\propto ~c^{r_1}~\frac{\partial}{\partial x^i}~,\snk\\[5pt]
K_i ~\propto ~c^{s_0}\le(x_i\,\frac{1}{c}\,\frac{\partial}{\partial t}+c\,t\,\frac{\partial}{\partial
x^i}\ri)=c^{s_0-1}~x_i\,\frac{\partial}{\partial t}+~c^{s_0+1}~t\,\frac{\partial}{\partial x^i}~,\snk
\eeak
one should choose $r_0=1$ because the Hamiltonian must exist in both cases, and $s_0=1$ for Carroll or $s_0=-1$ for Galilei. These restrictions, combined with \eqref{CG}, determine all the remaining exponents uniquely, namely, $r_1=0$ for both cases, and $k={1}/{2}$ for Carroll or $k=0$ for Galilei.

\vspace{.1cm}

As a result, we find the super-Carroll and super-Galilei algebras. The super-Carroll algebra is generated by:
\beak{lll}\phantomsection\label{scgenerators}
K^{^{(\rm{C})}}_i = c~ \mathcal{J}_{0i}~,~\quad & P^{^{(\rm{C})}}_0 = c~ \mathcal{P}_0~,~\quad & Q^{^{(\rm{C})}}_{\a} = \sqrt{c}~ \mathcal{Q}_{\a}~,\\[8pt]
J^{^{(\rm{C})}}_{ij} = \mathcal{J}_{ij}~,~\quad & P^{^{(\rm{C})}}_{i} = \mathcal{P}_{i}~,~\quad & \bar{Q}^{^{(\rm{C})}}_{\dot\alpha} = \sqrt{c}~
\mathcal{\bar{Q}}_{\dot\alpha}~,\nk
\eeak
which have the following nonzero (anti)commutators
\beak{lll}\phantomsection\label{SCA}
[K^{^{(\rm{C})}}_i,J^{^{(\rm{C})}}_{jk}] = -\,i~\d_{i[j}~K^{^{(\rm{C})}}_{k]}~,~~~~&
[J^{^{(\rm{C})}}_{ij},J^{^{(\rm{C})}}_{kl}] = i~(\d_{[i|k|}J^{^{(\rm{C})}}_{j]l} - \d_{[i|l|}J^{^{(\rm{C})}}_{j]k})~,~~~~&
[K^{^{(\rm{C})}}_i,P^{^{(\rm{C})}}_j] = -\,i\,\d_{ij}~P^{^{(\rm{C})}}_0~,~\label{sCa}\\[8pt]
{}[J^{^{(\rm{C})}}_{ij},P^{^{(\rm{C})}}_k] = i~\d_{[i|k|}P^{^{(\rm{C})}}_{j]}~,~~~~&
[J^{^{(\rm{C})}}_{ij},Q^{^{(\rm{C})}}_\a] = (\s_{ij})_{\a}{}^{\b}~Q^{^{(\rm{C})}}_\b~,~~~~&
[J^{^{(\rm{C})}}_{ij},\bar{Q}^{^{(\rm{C})}}_{\dot\a}] = (\bar{\s}_{ij})_{\dot\a}{}^{\dot\b}~\bar{Q}^{^{(\rm{C})}}_{\dot\b}~,~\\[8pt]
\{Q^{^{(\rm{C})}}_\a,\bar{Q}^{^{(\rm{C})}}_{\dot\a}\} = -~(\s^0)_{\a\dot\a}~P^{^{(\rm{C})}}_0. && \nk
\eeak
This algebra match precisely the 
$\mathcal{N}$=1 subalgebra of the Carroll superconformal
algebra constructed
in \cite{Bagchi:2022owq} and also agrees with the flat limit of the
AdS super-Carroll algebra discussed in \cite{Bergshoeff:2015wma}.

\vspace{.10cm}

The super-Galilei algebra is generated by:
\beak{lll}\phantomsection\label{sggenerators}
K^{^{(\rm{G})}}_i = \frac{1}{c}~ \mathcal{J}_{0i}~,~\quad & P^{^{(\rm{G})}}_0 = c~ \mathcal{P}_0~,~\quad & Q^{^{(\rm{G})}}_{\a} = \mathcal{Q}_{\a}~,\\[8pt]
J^{^{(\rm{G})}}_{ij} = \mathcal{J}_{ij}~,~\quad & P^{^{(\rm{G})}}_{i} = \mathcal{P}_{i}~,~\quad & \bar{Q}^{^{(\rm{G})}}_{\dot\alpha} =
\mathcal{\bar{Q}}_{\dot\alpha}~,\nk
\eeak
which have the following nonzero commutators
\beak{lll}\phantomsection\label{SGA}
[K^{^{(\rm{G})}}_i,J^{^{(\rm{G})}}_{jk}] = -\,i~\d_{i[j}~K^{^{(\rm{G})}}_{k]}~,~~~~&
[J^{^{(\rm{G})}}_{ij},J^{^{(\rm{G})}}_{kl}] = i~(\d_{[i|k|}J^{^{(\rm{G})}}_{j]l} - \d_{[i|l|}J^{^{(\rm{G})}}_{j]k})~,~~~~&
[K^{^{(\rm{G})}}_i,P^{^{(\rm{G})}}_0] = -\,i\,P^{^{(\rm{G})}}_i~,\label{sGa}\\[8pt]
{}[J^{^{(\rm{G})}}_{ij},P^{^{(\rm{G})}}_k] = i~\d_{[i|k|}P^{^{(\rm{G})}}_{j]}~,~~~~&
[J^{^{(\rm{G})}}_{ij},Q^{^{(\rm{G})}}_\a] = (\s_{ij})_{\a}{}^{\b}~Q^{^{(\rm{G})}}_\b~,~~~~&
[J^{^{(\rm{G})}}_{ij},\bar{Q}^{^{(\rm{G})}}_{\dot\a}] = (\bar{\s}_{ij})_{\dot\a}{}^{\dot\b}~\bar{Q}^{^{(\rm{G})}}_{\dot\b}~,\\[8pt]
\{Q^{^{(\rm{G})}}_\a,\bar{Q}^{^{(\rm{G})}}_{\dot\a}\} = -~(\s^i)_{\a\dot\a}~P^{^{(\rm{G})}}_i~. & &\nk
\eeak
For each one of these algebras we can introduce an appropriate coordinate system $\{t,~x^i,~\theta^{\a},~\bar{\theta}^{\dot
\a}\}$ via the coset manifold approach.
These coordinates parametrize a choice of a representative $\Omega(t,x^i,\theta^{\a},\bar{\theta}^{\dot\a})$ for every left
coset generated by the orbits of right multiplication with the subgroup elements generated by spatial rotations and boosts.

\vspace{.2cm}

A particular, but not often practical, representation of $\Omega(t,x^i,\theta^{\a},\bar{\theta}^{\dot\a})$
is the exponential map
\beak{l}
\Omega(t,x^i,\theta^{\a},\bar{\theta}^{\dot\a})=~e^{i(-\,tP_0\,-\,x^iP_i\,+\,\theta^\a Q_\a\,+\,\bar{\theta}^{\dot\a}\bar{Q}_{\dot\a})}\nk~.
\eeak
We can use this representation to deduce the transformation of these coordinates under the action of the above symmetry
generators as a result of a motion on the coset space $\Omega(t,x^i,\theta^{\a},\bar{\theta}^{\dot\a})\to \Omega(t',x'^i,
\theta'^{\a},\bar{\theta}'^{\dot\a})$ produced by left group multiplication of any group element $g$ with the coset
representative $\Omega$:
\beak{l}
g~\Omega(t,x^i,\theta^{\a},\bar{\theta}^{\dot\a})~=~\Omega(t',x'^i,\theta'^{\a},\bar{\theta}'^{\dot\a})~h(t,x^i,\theta^{\a},
\bar{\theta}^{\dot\a},g)\nk~,
\eeak
where $h(t,x^i,\theta^{\a},\bar{\theta}^{\dot\a},g)$ is a subgroup element generated by particular spatial rotations and boosts, which depend on the coordinates and the group element $g$:
\beak{l}
h(t,x^i,\theta^{\a},\bar{\theta}^{\dot\a},g)=~e^{i\,W^i(t,\,x^i,\,\theta^{\a},\,\bar{\theta}^{\dot\a},\,g)K_i}~e^{i\,W^{ij}(t,\,x^i,\,
\theta^{\a},\,\bar{\theta}^{\dot\a},\,g)J_{ij}}\nk~.
\eeak

We find that under Carrollian supersymmetry ($\d^{^{(\rm{C})}}_{S}$) and Carrollian boosts ($\d^{^{(\rm{C})}}_{B}$)
the coordinates transformations are:
\beak{ll}\nk
\d^{^{(\rm{C})}}_{S}t = -\,\frac{i}{2}~\Big(\epsilon^{\a}(\s^0)_{\a\dot\a}\bar{\theta}^{\dot\a}~-~\theta^{\a}(\s^0)_{\a\dot\a}\bar{\epsilon}^{\dot\a}\Big) ~,~\quad\qquad& \d^{^{(\rm{C})}}_{B}t = b^ix_i ~,~\snk\\[8pt]
\d^{^{(\rm{C})}}_{S}x^i = 0 ~,~\qquad& \d^{^{(\rm{C})}}_{B}x^i = 0 ~,~\snk\\[8pt]
\d^{^{(\rm{C})}}_{S}\theta^{\a} = \epsilon^\a ~,~\qquad& \d^{^{(\rm{C})}}_{B}\theta^{\a} = 0 ~,~\snk\\[8pt]
\d^{^{(\rm{C})}}_{S}\bar{\theta}^{\dot\a} = \bar{\epsilon}^{\dot\a} ~,~\quad\qquad& \d^{^{(\rm{C})}}_{B}\bar{\theta}^{\dot\a} = 0 ~.~\snk
\eeak
Additionally, under Galilean supersymmetry ($\d^{^{(\rm{G})}}_{S}$) and Galilean boosts ($\d^{^{(\rm{G})}}_{B}$)
the coordinates transform as follows:
\beak{ll}\nk
\d^{^{(\rm{G})}}_{S}t = 0 ~,~\qquad& \d^{^{(\rm{G})}}_{B}t = 0 ~,~\snk\\[8pt]
\d^{^{(\rm{G})}}_{S}x^i = -\,\frac{i}{2}~\Big(\epsilon^{\a}(\s^i)_{\a\dot\a}\bar{\theta}^{\dot\a}~-~\theta^{\a}(\s^i)_{\a\dot\a}\bar{\epsilon}^{\dot\a}\Big) ~,~\quad\qquad& \d^{^{(\rm{G})}}_{B}x^i = b^it ~,~\snk\\[8pt]
\d^{^{(\rm{G})}}_{S}\theta^{\a} = \epsilon^\a ~,~\qquad& \d^{^{(\rm{G})}}_{B}\theta^{\a} = 0 ~,~\snk\\[8pt]
\d^{^{(\rm{G})}}_{S}\bar{\theta}^{\dot\a} = \bar{\epsilon}^{\dot\a} ~,~\quad\qquad& \d^{^{(\rm{G})}}_{B}\bar{\theta}^{\dot\a} = 0 ~.~\snk
\eeak

A useful observation is that both Carrollian and Galilean boost transformations of the coordinates are nilpotent:
\beak{c}\nk\phantomsection\label{BN}
\text{Carrollian:}~~~t~~\xrightarrow{\d^{^{(\rm{C})}}_{B}} ~x^i~ \xrightarrow{\d^{^{(\rm{C})}}_{B}} ~0~~,\snk\\[3mm]
\text{Galilean:}~~~~\,\,x^i~\xrightarrow{\d^{^{(\rm{G})}}_{B}} ~~t\, ~\xrightarrow{\d^{^{(\rm{G})}}_{B}} ~0~~.\snk
\eeak
Moreover, we define fields and superfields on the appropriate (super)space and time manifolds. The transformation of these (super)fields is defined based on the particular irreducible representation of the fields and the above coordinate transformations. The infinitesimal Carrollian/Galilean supersymmetry and boost
transformations\footnote{By $\bar{\d}$ we denote the difference between the transformed (super)field and the original
(super)field at the same position: $\bar{\d}\Phi\coloneqq~\Phi'(z)-\Phi(z)$. This is in contrast to $\d\Phi$ which denotes
the difference between the transformed (super)field evaluated in the new position minus the original (super)field evaluated
in the original position: $\d\Phi\coloneqq~\Phi'(z')-\Phi(z)$.} of superfields $\Phi(t,x^i,\theta^{\a},
\bar{\theta}^{\dot\a})$ are:
\beak{ll}\nk
\bar{\d}^{^{(\rm{C})}}_{S}\Phi\coloneqq~i\,\big(\epsilon^{\a}\mathbf{Q}^{^{(\rm{C})}}_{\a}+\bar{\epsilon}^{\a}\bar{\mathbf{Q}}^{^{(\rm{C})}}_{\dot\a}\big)~,~\qquad\qquad\qquad& \bar{\d}^{^{(\rm{C})}}_{B}\Phi\coloneqq i\,b^i\,\mathbf{K}^{^{(\rm{C})}}_i\Phi~,\snk\\[10pt]
\bar{\d}^{^{(\rm{G})}}_{S}\Phi\coloneqq~i\,\big(\epsilon^{\a}\mathbf{Q}^{^{(\rm{G})}}_{\a}+\bar{\epsilon}^{\a}\bar{\mathbf{Q}}^{^{(\rm{G})}}_{\dot\a}\big)~,~\qquad\qquad\qquad& \bar{\d}^{^{(\rm{G})}}_{B}\Phi\coloneqq i\,b^i\,\mathbf{K}^{^{(\rm{G})}}_i\Phi~,\snk
\eeak
where the corresponding $\mathbf{Q}_{\a},~\bar{\mathbf{Q}}_{\dot\a},~\mathbf{K}_i$
differential operators are defined as follows:
\beak{ll}\nk\label{dro}
\mathbf{Q}^{^{(\rm{C})}}_{\a}=i~\frac{\partial}{\partial\theta^\a}+\tfrac{1}{2}~\bar{\theta}^{\dot\a}(\s^0)_{\a\dot\a}~\frac{\partial}{\partial t}~,~~~~~\qquad\qquad&\mathbf{Q}^{^{(\rm{G})}}_{\a}=i~\frac{\partial}{\partial\theta^\a}+\tfrac{1}{2}~\bar{\theta}^{\dot\a}(\s^i)_{\a\dot\a}~\frac{\partial}{\partial x^i}~,~\snk\\[3mm]
\bar{\mathbf{Q}}^{^{(\rm{C})}}_{\dot\a}=i~\frac{\partial}{\partial\bar{\theta}^{\dot\a}}+\tfrac{1}{2}~\theta^{\a}(\s^0)_{\a\dot\a}~\frac{\partial}{\partial t}~,~~~~~\qquad\qquad&\bar{\mathbf{Q}}^{^{(\rm{G})}}_{\dot\a}=i~\frac{\partial}{\partial\bar{\theta}^{\dot\a}}+\tfrac{1}{2}~\theta^{\a}(\s^i)_{\a\dot\a}~\frac{\partial}{\partial x^i}~,~\snk\\[3mm]
\mathbf{K}^{^{(\rm{C})}}_i=i~x_i\,\frac{\partial}{\partial t}
~,~\quad&\mathbf{K}^{^{(\rm{G})}}_i=i~t\,\frac{\partial}{\partial x^i} ~.~\snk
\eeak
It is straightforward to check that these differential operators together with time and space translations, $\mathbf{P}_0=-\,i~\frac{\partial}{\partial t}$ and $\mathbf{P}_i=-\,i~\frac{\partial}{\partial x^i}$, satisfy the super-Carroll \eqref{sCa} and super-Galilei \eqref{sGa} algebras respectively.

\vspace{.3cm}

Later in the paper we will consider theories with manifest Carrollian or Galilean supersymmetry. For that purpose, it is convenient to construct the appropriate superspace covariant derivatives. Their defining property is to impose constraints
that remain invariant under the corresponding supersymmetry transformations. Therefore, we have the Carrollian superspace covariant derivatives
\beak{l}\nk\phantomsection\label{ccderivative}
\D^{^{(\rm{C})}}_{\a}\coloneqq \frac{\partial}{\partial\theta^\a}+\frac{i}{2}~\bar{\theta}^{\dot\a}(\s^0)_{\a\dot\a}~\frac{\partial}{\partial t}~~,~~\qquad\qquad\Dd^{^{(\rm{C})}}_{\dot\a}\coloneqq \frac{\partial}{\partial\bar{\theta}^{\dot\a}}+\frac{i}{2}~\theta^{\a}(\s^0)_{\a\dot\a}~\frac{\partial}{\partial t}~,\nk\label{CD}\\[10pt]
\{\D^{^{(\rm{C})}}_{\a},\mathbf{Q}^{^{(\rm{C})}}_{\b}\}=\{\D^{^{(\rm{C})}}_{\a},\bar{\mathbf{Q}}^{^{(\rm{C})}}_{\dot\b}\}=0~,~\qquad\qquad\{\Dd^{^{(\rm{C})}}_{\dot\a},\mathbf{Q}^{^{(\rm{C})}}_{\b}\}=\{\Dd^{^{(\rm{C})}}_{\dot\a},\bar{\mathbf{Q}}^{^{(\rm{C})}}_{\dot\b}\}=0~,~\\[10pt]
\{\D^{^{(\rm{C})}}_{\a},\Dd^{^{(\rm{C})}}_{\dot\a}\}=~i\,(\s^0)_{\a\dot\a}~\partial_t~,
\eeak
and the Galilean superspace covariant derivatives
\beak{l}\nk\phantomsection\label{gcderivative}
\D^{^{(\rm{G})}}_{\a}\coloneqq \frac{\partial}{\partial\theta^\a}+\frac{i}{2}~\bar{\theta}^{\dot\a}(\s^i)_{\a\dot\a}~\frac{\partial}{\partial x^i}~~,~~\qquad\qquad\Dd^{^{(\rm{G})}}_{\dot\a}\coloneqq \frac{\partial}{\partial\bar{\theta}^{\dot\a}}+\frac{i}{2}~\theta^{\a}(\s^i)_{\a\dot\a}~\frac{\partial}{\partial x^i}~,\nk\label{GD}\\[10pt]
\{\D^{^{(\rm{G})}}_{\a},\mathbf{Q}^{^{(\rm{G})}}_{\b}\}=\{\D^{^{(\rm{G})}}_{\a},\bar{\mathbf{Q}}^{^{(\rm{G})}}_{\dot\b}\}=0~,~\qquad\qquad\{\Dd^{^{(\rm{G})}}_{\dot\a},\mathbf{Q}^{^{(\rm{G})}}_{\b}\}=\{\Dd^{^{(\rm{G})}}_{\dot\a},\bar{\mathbf{Q}}^{^{(\rm{G})}}_{\dot\b}\}=0~,~\\[10pt]
\{\D^{^{(\rm{G})}}_{\a},\Dd^{^{(\rm{G})}}_{\dot\a}\}=~i\,(\s^i)_{\a\dot\a}~\partial_i~.
\eeak

By comparing equations \eqref{SCA} and \eqref{SGA},
one can observe that the distinction between the two superalgebras lies in the
(anti)commutators of the corresponding generators $K_i,\,P_i,\,P_0,\,Q_\a,
\,\bar{Q}_{\ad}$. Nevertheless, the nature of the disparity suggests the existence of
a formal mapping between the two structures, achieved by interchanging temporal and
spatial directions ($t\longleftrightarrow \vec{x}$) or more specifically by
interchanging indices $0$ and $i$ ($0\longleftrightarrow i$). This formal relation,
which becomes exact in 1+1 dimensions, was initially introduced in \cite{Duval:2014uoa}
and remains valid for the supersymmetric extensions of these algebras, as it maps
$\sigma^0\,P_0\longleftrightarrow \sigma^i\, P_i$.

\vspace{.3cm}

Furthermore, this duality, rooted in a particle perspective that discerns between time and spatial
directions, can be extended to a broader p-brane viewpoint \cite{Bergshoeff:2020xhv}. In this broader
context, time is part of the longitudinal directions, which are treated distinctively from the
transverse directions to the brane. In $D=d+1$ dimensions one can
introduce an index $A=0,1,\dots,p$ which captures longitudinal directions and corresponds to a p-brane
where index $i=p+1,\dots,d$ refers to the remaining transverse directions which correspond to a
$(D-p-2)$-brane. The two superalgebras can be immediately generalized in the following manner:\\[3mm]
\underline{($D$\,$-$\,$p$\,$-$\,$2$)-brane super-Carroll algebra:}
\beak{lll}\phantomsection\label{dp2braneSCA}
[K^{^{(\rm{C})}}_{Ai},J^{^{(\rm{C})}}_{jk}] = -\,i~\d_{i[j}~K^{^{(\rm{C})}}_{|A|k]}~,~&
[J^{^{(\rm{C})}}_{ij},J^{^{(\rm{C})}}_{kl}] = i~(\d_{[i|k|}J^{^{(\rm{C})}}_{j]l} - \d_{[i|l|}J^{^{(\rm{C})}}_{j]k})~,~&
[K^{^{(\rm{C})}}_{Ai},P^{^{(\rm{C})}}_j] = -\,i\,\d_{ij}~P^{^{(\rm{C})}}_A,\\[8pt]
[K^{^{(\rm{C})}}_{Ai},J^{^{(\rm{C})}}_{BC}] = -\,i~\eta_{A[B}~K^{^{(\rm{C})}}_{|C|i]}~,~&
[J^{^{(\rm{C})}}_{AB},J^{^{(\rm{C})}}_{CD}] = i~(\eta_{[A|C|}J^{^{(\rm{C})}}_{B]D} - \eta_{[A|D|}J^{^{(\rm{C})}}_{B]C})~,~&
{}[J^{^{(\rm{C})}}_{ij},P^{^{(\rm{C})}}_k] = i~\d_{[i|k|}P^{^{(\rm{C})}}_{j]}~,\\[8pt]
{}[J^{^{(\rm{C})}}_{AB},P^{^{(\rm{C})}}_C] = i~\eta_{[A|C|}P^{^{(\rm{C})}}_{B]}~,&
[J^{^{(\rm{C})}}_{AB},Q^{^{(\rm{C})}}_\a] = (\g_{AB})_{\a}{}^{\b}~Q^{^{(\rm{C})}}_\b~,&
[J^{^{(\rm{C})}}_{ij},Q^{^{(\rm{C})}}_\a] = (\g_{ij})_{\a}{}^{\b}~Q^{^{(\rm{C})}}_\b,\\[8pt]
\{Q^{^{(\rm{C})}}_\a,Q^{^{(\rm{C})}}_{\b}\} = -~(\g^A)_{\a\b}~P^{^{(\rm{C})}}_A. &&\nk
\eeak
%
\underline{$p$\,-brane super-Galilei algebra:}
\beak{lll}\phantomsection\label{pbraneSGA}
[K^{^{(\rm{G})}}_{Ai},J^{^{(\rm{G})}}_{jk}] = -\,i~\d_{i[j}~K^{^{(\rm{G})}}_{|A|k]}~,~&
[J^{^{(\rm{G})}}_{ij},J^{^{(\rm{G})}}_{kl}] = i~(\d_{[i|k|}J^{^{(\rm{G})}}_{j]l} - \d_{[i|l|}J^{^{(\rm{G})}}_{j]k})~,~&
[K^{^{(\rm{G})}}_{Ai},P^{^{(\rm{G})}}_B] = i\,\eta_{AB}\,P^{^{(\rm{G})}}_i,\\[8pt]
[K^{^{(\rm{G})}}_{Ai},J^{^{(\rm{G})}}_{BC}] = -\,i~\eta_{A[B}~K^{^{(\rm{G})}}_{|C|i]}~,~&
[J^{^{(\rm{G})}}_{AB},J^{^{(\rm{G})}}_{CD}] = i~(\eta_{[A|C|}J^{^{(\rm{G})}}_{B]D} - \eta_{[A|D|}J^{^{(\rm{G})}}_{B]C}),~&
{}[J^{^{(\rm{G})}}_{ij},P^{^{(\rm{G})}}_k] = i~\d_{[i|k|}P^{^{(\rm{G})}}_{j]},\\[8pt]
{}[J^{^{(\rm{G})}}_{AB},P^{^{(\rm{G})}}_C] = i~\eta_{[A|C|}P^{^{(\rm{G})}}_{B]}~&
[J^{^{(\rm{G})}}_{AB},Q^{^{(\rm{G})}}_\a] = (\g_{ij})_{\a}{}^{\b}~Q^{^{(\rm{G})}}_\b~,~&
[J^{^{(\rm{G})}}_{ij},Q^{^{(\rm{G})}}_\a] = (\g_{ij})_{\a}{}^{\b}~Q^{^{(\rm{G})}}_\b,\\[8pt]
\{Q^{^{(\rm{G})}}_\a,Q^{^{(\rm{G})}}_{\b}\} = -~(\g^i)_{\a\b}~P^{^{(\rm{G})}}_i~. & &\nk
\eeak
It is now clear that there is a formal relation between these superalgebras generated by the
interchanges of indices $A$ and $i$: $A\longleftrightarrow i$. In this relation $P_A \longleftrightarrow P_i$ and $\g^A\,P_A \longleftrightarrow \g^i\,P_i$.

\section{Carroll fermions}\label{Cfermions}
Naturally, one would like to consider field theories that respect the above superalgebras.  Focusing on the bosonic
subalgebras (Carroll or Galilei), several methodologies have been developed in order to explore non-Lorentzian models.
Prototypical examples are the electric or magnetic Carroll (Galilei) boost invariant scalar field theories constructed by
various approaches such as the Hamiltonian formulation \cite{Henneaux:2021yzg}, the field expansion approach \cite{deBoer:2021jej}, the Lagrange multiplier method \cite{deBoer:2021jej}, and the seed Lagrangian method
\cite{Bergshoeff:2022qkx} (a review of each framework is provided in appendix \ref{Carroll scalar rev}).
Motivated by these techniques, except the
Lagrange multiplier method which fails for fermions due to the nonquadratic kinetic
term, we employ them to construct fermionic theories that exhibit Carroll (Galilei) boost invariance.

\subsection{Hamiltonian method}

The method of Hamiltonian formulation was first used to build Carroll scalar field theories \cite{Henneaux:2021yzg} (for a brief review see \ref{hf}). This method depends on having two independent variables in the formulation, the field and its conjugate momentum, that enable us to take different limits. We apply this method to construct Carroll Dirac Lagrangians. For this purpose, we start from the Lagrangian density of the massless Dirac field in four-dimensional Minkowski spacetime, which involves the speed of light $c$, as follows (up to total derivatives):
\begin{align}
    \mathcal{L}=-\,\bar\psi\,\g^\m\p_\m\,\psi&=-\,\frac{1}{2}\,\Big\{ \bar\psi\,\g^\m\p_\m\,\psi -(\p_\m\bar\psi)\,\g^\m\psi \Big\} \nonumber\\[8pt]
    &=-\,\frac{1}{2}\,\Big\{ \bar\psi\,\Big[~\frac{1}{c}\,\g^0\,\p_t+\g^i\p_i~\Big]\,\psi -\,\frac{1}{c}\,(\p_t\bar\psi)\,\g^0\,\psi-(\p_i\bar\psi)\,\g^i\psi ~ \Big\}\,,\label{diracLag}
\end{align}
where $\bar\psi:=\psi^\dagger i \g^0$ is the Dirac adjoint.
The canonical momenta conjugate to $\psi$ and $\bar{\psi}$ are
\be
\pi_{_\psi}=\frac{\p\mathcal{L}}{\p\dot{\psi}}=-\,\frac{1}{2c}\,\bar\psi\,\g^0\,,\qquad\qquad
\bar\pi_{_{\bar\psi}}=\frac{\p\mathcal{L}}{\p\dot{\bar\psi}}=\frac{1}{2c}\,\g^0\,\psi\,,\label{conjugate f}
\ee
and, as a result, the Hamiltonian density becomes
\be
\mathcal{H}\,=\,\pi_{_\psi}\,\dot{\psi}+\dot{\bar\psi}\,\bar\pi_{_{\bar\psi}}-\mathcal{L}\,=\,\bar\psi\,\g^i\p_i\,\psi\,.
\label{Hamilto Dira}
\ee
We can now derive the action of the massless Dirac field in the Hamiltonian formulation by\footnote{We note that the first two terms in \eqref{Dirac hamil} are the complex conjugates of each other, as expected, i.e. $(\pi_{_\psi}\,\dot{\psi})^\dagger=\dot{\bar\psi}\,\bar\pi_{_{\bar\psi}}$.}
\begin{align}
S\,[\psi,\bar\psi,\pi_{_\psi},\bar\pi_{_{\bar\psi}}]
=\int\,dt\,d^3x\,\Big\{\,\pi_{_\psi}\,\dot{\psi}+\dot{\bar\psi}\,\bar\pi_{_{\bar\psi}}-\mathcal{H}\,\Big\}
= \int\,dt\,d^3x\,\le\{ ~\pi_{_\psi}\,\dot{\psi}+\dot{\bar\psi}\,\bar\pi_{_{\bar\psi}}-\bar\psi\,\g^i\p_i\,\psi ~\ri\}\,.\label{Dirac hamil}
\end{align}
Since this formulation includes independent variables, it is possible to take the Carroll limit $c\rightarrow 0$ in two different ways. First, we may take the limit explicitly without rescaling the fields. In this case, the action \eqref{Dirac hamil} does not involve any explicit $c$ dependence, so we simply replace the fields $\pi_\psi$ and $\bar\pi_{_{\bar\psi}}$ by auxiliary fields $\pi_\eta$ and $\bar\pi_{_{\bar\eta}}$, respectively, to indicate that they lose their canonical relation \eqref{conjugate f} with $\psi$ and $\bar{\psi}$ after taking the limit. By doing this, we obtain the so-called ``magnetic'' Carroll Dirac action in the Hamiltonian formulation as follows:
\begin{align}
S_\mathrm{mC}=\int\,dt\,d^3x\,\Big\{ ~\pi_{_\eta}\,\dot{\psi}\,+\,\dot{\bar\psi}\,\bar\pi_{_{\bar\eta}}\,-\,\bar\psi\,\g^i\p_i\,\psi ~\Big\}\,.
\label{magnetic Dirac}
\end{align}
Alternatively, we can take the limit after rescaling the fields in \eqref{Dirac hamil}, i.e. $\pi_\psi\rightarrow\frac{1}{\sqrt{c}}\,\pi_\psi'$ and $\psi\rightarrow{\sqrt{c}}\,\psi'$ (similarly $\bar\pi_{_{\bar\psi}}\rightarrow\frac{1}{\sqrt{c}}\,\bar\pi_{_{\bar\psi}}'$ and $\bar\psi\rightarrow{\sqrt{c}}\,\bar\psi'$) preserving the canonical structure. This effectively eliminates $c$ from the canonical relations \eqref{conjugate f} and ensures their well-definedness in the limit. Therefore, by rescaling the fields in the action \eqref{Dirac hamil}, taking the limit $c\rightarrow 0$, and dropping the prime, we find the so-called ``electric'' Carroll Dirac action in the Hamiltonian formulation
\be
S_\mathrm{eC}=\int\,dt\,d^3x\,\le\{ ~\pi_{_\psi}\,\dot{\psi}\,+\,\dot{\bar\psi}\,\bar\pi_{_{\bar\psi}} ~\ri\}\,.\label{electric Dirac}
\ee

\subsection{Field expansion method}

By expanding the field around $c=0$, the field expansion method was initially employed to build Carroll scalar Lagrangians \cite{deBoer:2021jej} (see \ref{fe} for a short overview). The expansion results in extra field that appears in the Lagrangian and leads to obtaining electric and magnetic Carroll invariant Lagrangians at the same time. We use this method for spin $1/2$ fermions. For this purpose, let us consider again the Lagrangian density of the massless Dirac field in four dimensions
\be
\mathscr{L}=-\,\bar{\varPsi}\,(\gamma^\m\p_\m)\,\varPsi=-\,\bar{\varPsi}\,\Big(\,\frac{1}{c}\,\g^0\p_t+\g^i\p_i\,\Big)\,\varPsi\,. \label{Dirac Lag}
\ee
We then perform an expansion around $c=0$ in the Dirac field\footnote{Unlike the scalar field expansion \eqref{Phi expansion}, the field expansion for fermions has also odd powers of $c$ that contribute to the Carroll invariant terms.}
\be
{\varPsi}=c^\b\Big(\psi+c\,{\eta}+\mathcal{O}(c^2)\Big)\,, \label{Dirac expan}
\ee
for some $\b$, and introduce $\mathcal{L}_\mathrm{eC}$ and $\mathcal{L}_\mathrm{mC}$ through the Lagrangian density
\be
\mathscr{L}=c^{2\b-1}\Big( \mathcal{L}_\mathrm{eC}+c\,\mathcal{L}_\mathrm{mC}+\mathcal{O}(c^2)\Big)\,.\label{L expansion}
\ee
By substituting the field expansion \eqref{Dirac expan} into the Lagrangian \eqref{Dirac Lag} and comparing with \eqref{L expansion}, we can extract both the electric and magnetic Carroll Dirac Lagrangians, respectively, as follows:
\begin{align}
    \mathcal{L}_\mathrm{eC}&= -\,\bar{\psi}\,\g^0\p_t\,\psi\,,\label{elec f} \\[8pt]
    \mathcal{L}_\mathrm{mC}&=-\,\le(\, \bar{\psi}\,\g^i\p_i\psi\,+\,\bar{\eta}\,\g^0\p_t\,\psi\,+\,\bar{\psi}\,\g^0\p_t\,\eta\,\ri)\,.\label{magn f}
\end{align}


We next explore the transformation properties of these Lagrangians under Carroll boost. To do this, we use the fact that the spinor field $\psi$ transforms like a scalar \eqref{Carroll trans} under Carroll boosts. Indeed, as we know, when subjected to a Lorentz boost, the commutator between the Lorentz boost generator $\mathcal{J}_{0i}$ and a spinor field $\psi$ takes the form $[\,\mathcal{J}_{0i}\,,\,\psi\,]\propto\sigma_{0i}\,\psi$. By introducing the Carroll boost generator \eqref{scgenerators}, $K_i^{^{(\rm{C})}}=c\,\mathcal{J}_{0i}$, this commutation relation will take an overall factor of $c$ on the right-hand side\footnote{According to our convention \ref{conven}, $c$ is included in the coordinate component, so $\sigma_{0i}$ remains unchanged.}. As we consider the Carroll limit $c\rightarrow 0$, the commutator simplifies to $[\,K_i^{^{(\rm{C})}}\,,\,\psi\,]=0$, demonstrating that the Carroll spinor field transforms as a scalar under the Carroll boost. As a result of this observation, we find that the electric \eqref{elec f} and magnetic \eqref{magn f} Carroll Dirac Lagrangians remain invariant (up to total derivatives) when subjected to the following Carroll boost transformations
\begin{align}
\d^{^{(\rm{C})}}_{B}\psi={b}^i{x}_i\,\p_t\,\psi\,,\qquad\qquad
\d^{^{(\rm{C})}}_{B}\eta={b}^i{x}_i\,\p_t\,\eta\,+\,\frac{1}{2}\,\g^0\,\g^i \,b_i\,\psi\,, \label{Carroll Dirac trans}
\end{align}
where $b_i$ are the Carroll boost parameters. The next subsection will provide another approach to obtain these transformations.

\vspace{.3cm}

The equations of motion for the electric \eqref{elec f} and magnetic \eqref{magn f} Lagrangians read as follows:
\begin{align}
    \mathrm{eC}-e.o.m:& \qquad  \g^0\p_t\,\psi=0\,, \\[8pt]
    \mathrm{mC}-e.o.m:& \qquad  \g^0\p_t\,\psi=0\,,\qquad \g^0\p_t\,\eta+\g^i\p_i\,\psi=0\,. \label{fer}
\end{align}
Similar to the bosonic case \eqref{bosonic}, a matrix equation can be used to represent two magnetic Carroll Dirac field equations \eqref{fer} more compactly
\be
\qquad \mathrm{F}\,\mathrm{\Psi}=0\,, \quad\qquad \mbox{with}\quad \mathrm{F}:=\begin{pmatrix} \g^i\p_i & \g^0\p_t \\ \g^0\p_t & 0  \end{pmatrix}\,,\quad \mathrm{\Psi}:=\begin{pmatrix} \psi \\ \eta \end{pmatrix}\,.\label{fermionic oper}
\ee
In this form, the magnetic Carroll Dirac Lagrangian density \eqref{magn f} can be expressed by
\be
\mathcal{L}_\mathrm{mC}=-\,\mathrm{\bar\Psi}\,\mathrm{F}\,\mathrm{\Psi}\,. \label{Matrix fermion}
\ee

Finally, we note that the electric Carroll Lagrangian \eqref{elec f} is equivalent to the one in the Hamiltonian formulation \eqref{electric Dirac} by imposing the conditions $\pi_{_\psi}=-\,\frac{1}{2}\,\bar\psi\,\g^0$ and $\bar\pi_{_{\bar\psi}}=\frac{1}{2}\,\g^0\,\psi$, which follow from rescaling the fields in \eqref{conjugate f}. Moreover, the magnetic Carroll Lagrangian \eqref{magn f} would be equivalent to \eqref{magnetic Dirac} by defining $\pi_{\eta}:=-\,\bar\eta\,\g^0$ and $\bar\pi_{_{\bar\eta}}:=\g^0\,\eta$, which arises from the fact of being Hermitian.

\subsection{Seed Lagrangian method}
The seed Lagrangian method is an alternative approach of constructing Carroll invariant theories.
It was introduced in \cite{Bergshoeff:2022qkx} as a method to construct new non-Lorentzian scalar theories and it relies on the nilpotence of the Carrollian and Galilean boost transformations \eqref{BN}. We use this method as an alternative approach to constructing the magnetic Carroll Dirac Lagrangian (for a review of magnetic scalar Lagrangian refer to \ref{sl}). In this case the seed Lagrangian is the
magnetic Galilei Dirac Lagrangian \eqref{elec f2} (discussed in the next section):
\be
\mathcal{L}_\mathrm{mG} = -\, \bar{\psi}\,\g^i\p_i\psi\,.
\ee
Obviously, this theory is not invariant under Carrollian boosts, specifically its transformation under $\d^{^{(\rm{C})}}_B$ is
\beak{c}\nk\label{dCmG}
\d^{^{(\rm{C})}}_{B}S_\mathrm{mG}=\int dt\, d^3x~\Big\{\,\bar{\psi}\,\g^i b_i\,\p_t\psi\Big\}\,,
\eeak
where we used the fact that (i) the integral's measure is invariant under infinitesimal boost transformations, (ii)
the spinor field $\psi(t,x)$ transforms like a scalar under boosts\footnote{Notice that the commutator of $K_i$ with the spinorial supersymmetry charges $Q_{\a}$ and $\bar{Q}_{\dot\a}$ in both super Carroll \eqref{sCa} and super Galilei \eqref{sGa} algebras are zero. This can also be verified by the commutator of the differential operators $\mathbf{Q}_{\a},~\bar{\mathbf{Q}}_{\dot\a},~\mathbf{K}_i$ in \eqref{dro}.}
\beak{c}\nk
\d^{^{(\rm{C})}}_{B}\psi(t,x)=\psi'(t',x')-\psi(t,x)=0~~~~~~\Rightarrow~~~~~~\bar{\d}^{^{(\rm{C})}}_{B}\psi(t,x)=i\,b^i\mathbf{K}^{^{(\rm{C})}}_{i}\psi=-\,b^ix_i~\p_t\psi~,
\eeak
and (iii) the spatial derivative of $\psi$ transforms
\beak{c}
\d^{^{(\rm{C})}}_{B}\p_i\psi(t,x)=\p'_i\psi'(t',x')-\p_i\psi(t,x)=-\,b_i~\p_t\psi\,,~~~~~~~~~\Big([K^{^{(\rm{C})}}_i,P^{^{(\rm{C})}}_j] = -\,i\d_{ij}P^{^{(\rm{C})}}_0\Big)\nk\,.
\eeak
Now observe that the quantity $\p_t\psi$ is invariant under Carrollian boost transformations:
\beak{c}
\d^{^{(\rm{C})}}_{B}\p_t\psi(t,x)=\p'_t\psi'(t',x')-\p_t\psi(t,x)=0\,,~~~~~~~~~~~~\Big([K^{^{(\rm{C})}}_i,P^{^{(\rm{C})}}_0] = 0\Big)\nk\,.
\eeak
Hence the nilpotence of $\d^{^{(\rm{C})}}_{B}$ is realized in this theory in the following manner:
\beak{c}\nk
\p_i\psi~\xrightarrow{\d^{^{(\rm{C})}}_{B}} ~\p_t\psi~ \xrightarrow{\d^{^{(\rm{C})}}_{B}} ~0~.
\eeak
A consequence of this, is that the nonzero transformation \eqref{dCmG} of the seed Lagrangian can be compensated by the addition of a second term in the Lagrangian which involves a fermionic Lagrange multiplier $\eta$
\beak{c}\nk
\mathcal{L}^{\eta}_\mathrm{mG}=-\,\bar{\eta}\,\g^0\p_t\,\psi-\bar{\psi}\,\g^0\p_t\,\eta\,.
\eeak
By choosing the transformation of the Lagrange multiplier $\eta$ under Carrollian boosts appropriately
\beak{c}\nk
\d^{^{(\rm{C})}}_{B}\eta=-\,\frac{1}{2}\,\g^0\g^i\,b_i\,\psi~~~\Rightarrow~~~\bar{\d}^{^{(\rm{C})}}_{B}\eta=i\,b^i\,\mathbf{K}^{^{(\rm{C})}}_{i}\eta-\frac{1}{2}\,\g^0\g^i\,b_i\,\psi=-\,b^ix_i\,\p_t\,\eta-\frac{1}{2}\,\g^0\g^i\,b_i\,\psi\,,
\eeak
we can construct a Carrollian invariant theory for the Dirac field where the spatial derivatives dominate (magnetic)
\beak{c}\nk
\mathcal{L}_\mathrm{mC}=-\,\le(\, \bar{\psi}\,\g^i\p_i\psi+\bar{\eta}\,\g^0\p_t\,\psi
    +\bar{\psi}\,\g^0\p_t\,\eta\,\ri)\,.
\eeak
The Lagrange multiplier $\eta$ of the seed Lagrangian method we just described can be understood as the
$\eta$ fermionic field in the field expansion method \eqref{Dirac expan}.


\section{Galilei fermions}\label{Gfermions}

In this section, we use the methods that have been employed to build Carroll scalar fields \cite{Henneaux:2021yzg, deBoer:2021jej, Bergshoeff:2022qkx} to develop fermionic theories that respect Galilei boost symmetry\footnote{Galilei fermions are also discussed in \cite{Banerjee:2022uqj,Sharma:2023chs}}. We exclude the Lagrange multiplier method, which fails for fermions, and use other methods to derive electric and magnetic Galilei fermionic field Lagrangians. We refer the reader to appendix \ref{Gscalar} for a review of the Galilei scalar field theories obtained by different techniques.

\subsection{Hamiltonian method}
As we explained in the Galilei scalar part \ref{HmG}, the standard Hamiltonian formulation method for Carroll Lagrangians \cite{Henneaux:2021yzg} cannot be used to derive electric and magnetic Galilei Lagrangians. Instead, we adopt a novel approach called the ``Hamiltonian-like'' formulation, which involves introducing new conjugate momenta for the fields. To illustrate this, we consider the massless Dirac field in four-dimensional Minkowski spacetime, whose Lagrangian density is given by \eqref{diracLag}. We then define $\mathring{\psi}_i\equiv \p_i \psi$ and $\mathring{\bar\psi}_i\equiv \p_i \bar\psi$, and introduce the canonical momenta associated with the fields $\psi$ and $\bar\psi$ as follows:
\be
\Pi^i_{_\psi}=\frac{\p\mathcal{L}}{\p\mathring{\psi}_i}=-\,\frac{1}{2}\,\bar\psi\,\g^i\,,\qquad\qquad
\bar\Pi^i_{_{\bar\psi}}=\frac{\p\mathcal{L}}{\p\mathring{\bar\psi}_i}=\frac{1}{2}\,\g^i\,\psi\,.\label{conjugate f g}
\ee
Thus, the Hamiltonian-like density, which is different from the Hamiltonian density \eqref{Hamilto Dira}, becomes
\be
\mathscr{H}\,=\,\Pi^i_{_\psi}\,\mathring{\psi}_i\,+\,\mathring{\bar\psi}_i\,\bar\Pi^i_{_{\bar\psi}}\,-\,\mathcal{L}\,=\,\frac{1}{c}\,\bar\psi\,\g^0\p_t\,\psi\,.
\label{Hamilto Dira ga}
\ee
As a result, the action of the massless Dirac field in the Hamiltonian-like formulation reads\footnote{As expected, the first two terms in \eqref{Dirac hamil ga} are the complex conjugates of each other, i.e. $(\Pi^i_{_\psi}\,\mathring{\psi}_i)^\dagger=\mathring{\bar\psi}_i\,\bar\Pi^i_{_{\bar\psi}}$.}
\begin{align}
S
=\int\,dt\,d^3x\,\Big\{\,\Pi^i_{_\psi}\,\mathring{\psi}_i\,+\,\mathring{\bar\psi}_i\,\bar\Pi^i_{_{\bar\psi}}\,-\,\mathscr{H}\,\Big\}
= \int\,dt\,d^3x\,\le\{ ~\Pi^i_{_\psi}\,\mathring{\psi}_i\,+\,\mathring{\bar\psi}_i\,\bar\Pi^i_{_{\bar\psi}}\,-\,\frac{1}{c}\,\bar\psi\,\g^0\p_t\,\psi ~\ri\}\,.\label{Dirac hamil ga}
\end{align}
We have reached a point where we can apply the Galilei limit $c\rightarrow\infty$ to \eqref{Dirac hamil ga} in two distinct ways. First, we take the limit directly without rescaling the fields, which also avoids any singularity in \eqref{conjugate f g}. This implies that \eqref{Dirac hamil ga} results in the magnetic Galilei Dirac action
\be
S_\mathrm{mG}=\int\,dt\,d^3x\,\le\{ ~\Pi^i_{_\psi}\,\mathring{\psi}_i\,+\,\mathring{\bar\psi}_i\,\bar\Pi^i_{_{\bar\psi}} ~\ri\}\,.\label{mGDirac}
\ee
We can also apply the Galilei limit $c\rightarrow\infty$ to \eqref{Dirac hamil ga} after rescaling the fields by a factor of $\sqrt{c}$. That is, we replace $\Pi^i_\psi$ by $\frac{1}{\sqrt{c}}\,\Pi^i_\psi$ and $\psi$ by ${\sqrt{c}}\,\psi$, and similarly for $\bar\Pi^i_{_{\bar\psi}}$ and $\bar\psi$. The rescaling eliminates the factor of $\frac{1}{c}$ from \eqref{Dirac hamil ga}, but it also violates the canonical relations \eqref{conjugate f g} in the limit. Therefore, we have to replace $\Pi^i_{_{\psi}}$ and $\bar\Pi^i_{_{\bar\psi}}$ by new auxiliary fields $\Pi^i_{_{\eta}}$ and $\bar\Pi^i_{_{\bar\eta}}$ respectively after taking the limit. By doing this, we obtain the electric Galilei Dirac action
\begin{align}
S_\mathrm{eG}=\int\,dt\,d^3x\,\Big\{ ~\Pi^i_{_\eta}\,\mathring{\psi}_i\,+\,\mathring{\bar\psi}_i\,\bar\Pi^i_{_{\bar\eta}}\,-\,\bar\psi\,\g^0\p_t\,\psi ~\Big\}\,.
\label{eGDirac}
\end{align}

\subsection{Field expansion method}

The field expansion method for Carrollian fields \cite{deBoer:2021jej} uses $c=0$ as the expansion point. For Galilean fields, we choose $c=\infty$ as the expansion point. We applied this method to Galilei scalar fields in \ref{fG1} and reproduced Galilei scalar Lagrangians in this manner. Now we use this method for fermions. To do this, let us take into account the Lagrangian density of the massless Dirac field in four dimensions
\be
\mathscr{L}=-\,\bar{\varPsi}\,(\gamma^\m\p_\m)\,\varPsi=-\,\bar{\varPsi}\,\Big(\,\frac{1}{c}\,\g^0\p_t+\g^i\p_i\,\Big)\,\varPsi\,, \label{Dirac Lagra}
\ee
where $\bar{\varPsi}:=\varPsi^\dagger i\g^0$ is the Dirac adjoint. We can then make an expansion around $c=\infty$ in the Dirac field\footnote{Note that, unlike the scalar field expansion \eqref{Phi expansion2}, odd powers of $c$ are involved for fermions.}
\be
{\varPsi}=c^{-\b}\Big(\psi+c^{-1}\,{\eta}+\mathcal{O}(c^{-2})\Big)\,,\label{Galilei exp}
\ee
for some $\b$, and introduce $\mathcal{L}_\mathrm{mG}$ and $\mathcal{L}_\mathrm{eG}$ via the Lagrangian density
\be
\mathscr{L}=c^{-2\b}\Big( \mathcal{L}_\mathrm{mG}+c^{-1}\,\mathcal{L}_\mathrm{eG}+\mathcal{O}(c^{-2})\Big)\,. \label{L expan galilei}
\ee
When we use \eqref{Galilei exp} for the Lagrangian \eqref{Dirac Lagra} and compare it with \eqref{L expan galilei}, we can read both the magnetic and electric Galilei Dirac Lagrangians respectively
\begin{align}
    \mathcal{L}_\mathrm{mG}&= -\,\bar{\psi}\,(\,\g^i\p_i\,)\,\psi\,,\label{elec f2} \\[8pt]
    \mathcal{L}_\mathrm{eG}&=-\,\le(\, \bar{\psi}\,\g^0\p_t\psi\,+\,\bar{\eta}\,\g^i\p_i\,\psi\,+\,\bar{\psi}\,\g^i\p_i\,\eta\,\ri)\,.\label{magn f2}
\end{align}

We will now investigate the transformation properties of these Lagrangians under Galilei boost. We use the fact that the spinor field $\psi$ transforms like a scalar \eqref{Galilei trans} under Galilei boost. We know that a Lorentz boost on a spinor field $\psi$ makes the commutator $[\,\mathcal{J}_{0i}\,,\,\psi\,]\propto\sigma_{0i}\,\psi$. We define the Galilei boost generator \eqref{sggenerators} as $K^{^{(\rm{G})}}_i = \frac{1}{c}~ \mathcal{J}_{0i}$. Then the commutator gets a factor of $\frac{1}{c}$ on the right-hand side\footnote{We follow our convention \ref{conven} and include $c$ in the coordinate component, so $\sigma_{0i}$ stays the same.}. In the Galilei limit $c\rightarrow \infty$, the commutator becomes $[\,K_i^{^{(\rm{G})}}\,,\,\psi\,]=0$. This illustrates that the Galilei spinor field transforms as a scalar under Galilei boost. Because of this fact, we find that the magnetic \eqref{elec f2} and electric \eqref{magn f2} Galilei Dirac Lagrangians are invariant (up to total derivatives) under the following Galilei boost transformations:
\begin{align}
{\d}^{^{(\rm{G})}}_{B}\psi= -\,t\,b^i\,\p_i \,\psi\,,\qquad\qquad
{\d}^{^{(\rm{G})}}_{B}\eta=-\,t\,b^i\,\p_i \,\eta\,+\,\frac{1}{2}\,\g^i\,\g^0 \,b_i\,\psi\,, \label{Galilei Dirac trans}
\end{align}
where $b_i$ are the parameters of the Galilei boost. We will obtain these transformations in a different way in the next subsection.

\vspace{.3cm}

The magnetic Galilei Lagrangian \eqref{elec f2} gives the equation of motion $\g^i\p_i\,\psi=0$. The electric Galilei Lagrangian \eqref{magn f2} has the equation of motion in a compact form as
\be
\qquad \mathrm{F}\,\mathrm{\Psi}=0\,, \quad\qquad \mbox{with}\quad \mathrm{F}:=\begin{pmatrix} \g^0\p_t & \g^i\p_i \\ \g^i\p_i & 0  \end{pmatrix}\,,\quad \mathrm{\Psi}:=\begin{pmatrix} \psi \\ \eta \end{pmatrix}\,.\label{fermionic oper2}
\ee
In this form, the electric Galilei Dirac Lagrangian density \eqref{magn f2} can be expressed as
\be
\mathcal{L}_\mathrm{eG}=-\,\mathrm{\bar\Psi}\,\mathrm{F}\,\mathrm{\Psi}\,. \label{Matrix fermion2}
\ee

We remark that by imposing the canonical relations \eqref{conjugate f g}, the magnetic Galilei Dirac Lagrangian \eqref{elec f2} is the same as the one in the Hamiltonian formulation \eqref{mGDirac}. Furthermore, by defining $\Pi^i_{\eta}:=-\,\bar\eta\,\g^i$ and $\bar\Pi^i_{_{\bar\eta}}:=\g^i\,\eta$ the electric Galilei Dirac Lagrangian \eqref{magn f2} would be the same as \eqref{eGDirac}.

\subsection{Seed Lagrangian method}

The electric formulation of a Galilei boost invariant fermion can be also deduced via the seed Lagrangian method. In this case the seed Lagrangian is that of an electric Carroll fermion
\be
\mathcal{L}_\mathrm{eC} = -\, \bar{\psi}\,\g^0\p_t\psi\,.
\ee
This theory is not invariant under Galilean boost, however its transformation under $\d^{^{(\rm{G})}}_B$ takes the following form:
\beak{c}\nk\label{dGeC}
\d^{^{(\rm{G})}}_{B}S_\mathrm{eC}=\int dt\, d^3x~\Big\{\,\bar{\psi}\,\g^0 \,b^i\,\p_i\,\psi\Big\}\,.
\eeak
This is special because the transformation is the product of two factors and one of them ($\sim \pa_i\,\psi$) is Galilean boost invariant, hence the deformation of the theory can be absorbed by adding an appropriate compensating term. Equation \eqref{dGeC} can be easily verified by the transformation of the fermion and its derivatives under Galilean boosts:
\beak{l}\nk
\d^{^{(\rm{G})}}_{B}\psi(t,x)=\psi'(t',x')-\psi(t,x)=0~~~~~\Rightarrow~~~~~\bar{\d}^{^{(\rm{G})}}_{B}\psi(t,x)=i\,b^i\,\mathbf{K}^{^{(\rm{G})}}_{i}\psi=-\,t\,b^i\,\p_i\,\psi\,,\snk\\[8pt]
\d^{^{(\rm{G})}}_{B}\p_i\psi(t,x)=\p'_i\psi'(t',x')-\p_i\psi(t,x)=0\snk\,,\\[8pt]
\d^{^{(\rm{G})}}_{B}\p_t\psi(t,x)=\p'_t\psi'(t',x')-\p_t\psi(t,x)=-\,b^i\,\p_i\,\psi~.\snk
\eeak
In this theory, the nilpotence of $\d^{^{(\rm{G})}}_{B}$ is realized in the following manner:
\beak{c}\nk
\p_t\psi~\xrightarrow{\d^{^{(\rm{G})}}_{B}} ~\p_i\psi~ \xrightarrow{\d^{^{(\rm{G})}}_{B}} ~0~.
\eeak
To restore G-boost invariance, we must add the following compensating terms
\beak{c}\nk
\mathcal{L}^{\lambda}_\mathrm{eG}=-\,\bar{\lambda}^i\,\g^0\,\p_i\,\psi-\bar{\psi}\,\g^0\,\p_i\,\lambda^i\,.
\eeak
where $\lambda^i$ is a collection of compensating fields whose transformation under Galilean boosts are
\beak{c}\nk
\d^{^{(\rm{G})}}_{B}\lambda^i=\frac{1}{2}\,b^i\,\psi~~~~\Rightarrow~~~~\bar{\d}^{^{(\rm{G})}}_{B}\lambda^i=-\,t\,b^j\,\p_j\,\lambda^i+\frac{1}{2}\,b^i\,\psi\,.
\eeak
We conclude that the electric description of a Galilei boost invariant fermion can take the form
\beak{c}\nk\label{eG}
\mathcal{L}_\mathrm{eG}=-\,\le(\, \bar{\psi}\,\g^0\,\p_t\,\psi+\bar{\lambda}^i\,\g^0\,\p_i\,\psi
+\bar{\psi}\,\g^0\,\p_i\,\lambda^i\,\ri)\,.
\eeak
We note that the latter is equivalent to the electric Galilei Lagrangian in the Hamiltonian formulation \eqref{eGDirac}, by introducing $\Pi^i_\e:=-\,\bar\lambda^i\,\g^0$ and $\bar\Pi^i_{\bar\e}:=\g^0\,\lambda^i$. Moreover, it can be connected to the one in the field expansion method \eqref{magn f2}, by defining $\lambda^i:=-\,\g^0\,\g^i\,\e$.

\section{Carroll Boost Invariant Supersymmetric Theories}\label{sec5}
A natural question to ask is whether the various formulations (electric or magnetic) of Carroll or Galilei
fermions described in sections \ref{Cfermions} and \ref{Gfermions} can be combined with the
corresponding formulations of Carroll or Galilei scalars
\cite{Henneaux:2021yzg,deBoer:2021jej,Bergshoeff:2022qkx} (see appendices \ref{Carroll scalar rev} and \ref{Gscalar} for a
review) to define supersymmetric theories. As established in section \ref{super carroll algebra}, there
are two types of supersymmetries. We can have either (\emph{i}) \emph{C-supersymmetry} or
(\emph{ii}) \emph{G-supersymmetry}
\beak{c}\nk\label{nLsusy}
\texttt{C-supersymmetry}:~~ \{\,Q_{\a}\,,\bar{Q}_{\ad}\,\}=i (\s^0)_{\a\ad}~\p_{t}\snk\label{Csusy}~,\\[8pt]
\texttt{G-supersymmetry}:~~ \{\,Q_{\a}\,,\bar{Q}_{\ad}\,\}=i (\s^i)_{\a\ad}~\p_{i}\snk\label{Gsusy}~.
\eeak
In what follows, we demonstrate the construction of supersymmetric theories in which the Carrollian fermion can be incorporated into either a C-supermultiplet or a G-supermultiplet\footnote{Off-shell C/G supermultiplets are collections of bosonic and fermionic fields whose off-shell supersymmetry transformations satisfy the C/G supersymmetry algebras respectively, as defined in \eqref{nLsusy}.}, contingent upon whether we utilize its electric or magnetic description respectively. Similarly, magnetic Galilean fermions are embedded naturally in G-supermultiplets, but electric Galilean fermions are members of C-supersymmetric multiplets.

\vspace{.3cm}

The description of these theories is given in the appropriate superspace which makes the corresponding supersymmetries manifest. Using the C/G-supersymmetric covariant derivatives \eqref{CD} and \eqref{GD}, one can define irreducible representations of matter multiplets (including spin 0 and spin 1/2 fields) by imposing constraints on scalar superfields. In our description we will consider \emph{C-Chiral} and \emph{G-Chiral} superfields that are defined as follows:
\beak{l}\nk
\texttt{C-Chiral}:~~ \Dd^{^{(\rm{C})}}_{\ad}\Phi^{^{(\rm{C})}}=0~,~
\Phi^{^{(\rm{C})}}=\phi+\th^{\a}\psi_{a}+\th^2F+\tfrac{i}{2}\th^{\a}\thd^{\ad}(\s^0)_{\a\ad}~\p_t\phi-\tfrac{i}{2}\th^2\thd^{\ad}(\s^{0})^{\b}{}_{\ad}~\p_t\psi_{\b}\,,~~~~~~~\snk\label{Cchiral}\\[10pt]
\texttt{G-Chiral}:~~ \Dd^{^{(\rm{G})}}_{\ad}\Phi^{^{(\rm{G})}}=0~,~
\Phi^{^{(\rm{G})}}=\pi+\th^{\a}\lambda_{a}+\th^2G+\tfrac{i}{2}\th^{\a}\thd^{\ad}(\s^i)_{\a\ad}~\p_i\pi-\tfrac{i}{2}\th^2\thd^{\ad}(\s^{i})^{\b}{}_{\ad}~\p_i\lambda_{\b}~.~~~~~~~
\snk\label{Gchiral}\snk
\eeak

%

The superspace and component Lagrangian descriptions of these matter multiplets are respectively:
\beak{l}\nk\label{CGmatter}
\mathcal{L}_{\texttt{C-Chiral}}=~\bar{\Phi}^{^{(\rm{C})}}\Phi^{^{(\rm{C})}}~~~~~~~\rightarrow~~~~~~~L_{\texttt{C-WZ}}=-\,\bar{\phi}~\p_t^2\phi
\,+\,i~\bar{\psi}^{\ad}(\s^0)^{\a}{}_{\ad}~\p_t\,\psi_{\a}\,+\,\bar{F}F~~,~~~~~\snk\label{Cwz}\\[10pt]
\mathcal{L}_{\texttt{G-Chiral}}=~\bar{\Phi}^{^{(\rm{G})}}\Phi^{^{(\rm{G})}}~~~~~~~\rightarrow~~~~~~~L_{\texttt{G-WZ}}=\bar{\pi}~\p^i\p_i\pi
\,+\,i~\bar{\lambda}^{\ad}(\s^i)^{\a}{}_{\ad}~\p_i\,\lambda_{\a}\,
+\,\bar{G}G~~.~~~~~\snk\label{Gwz}
\eeak
We will refer to \eqref{Cwz} as the \emph{C-Wess-Zumino} model and, similarly, we will call \eqref{Gwz} the \emph{G-Wess-Zumino} model. It is evident that the C-Wess-Zumino model will be relevant for ``electric'' descriptions, whereas the G-Wess-Zumino model will be appropriate for ``magnetic'' descriptions.

\vspace{.3cm}

An interesting observation is that Carrollian supersymmetry \eqref{Csusy} is equivalent to the quantum
mechanical supersymmetry algebra, therefore one can use the techniques of \emph{Adinkras}
\cite{Faux:2004wb,Doran:2006it,Gates:2009me,Zhang:2011np} in order to classify and
construct representations of C-supersymmetry beyond the C-chiral \eqref{Cchiral} one.


\vspace{.3cm}

In this section, we study the transformation properties of the C-Wess-Zumino and G-Wess-Zumino models under Carrollian boosts, generated by $\mathbf{K}^{^{(\rm{C})}}_i$,  aiming toward the construction of Carrollian boost invariant supersymmetric theories. By calculating the commutator of $\mathbf{K}^{^{(\rm{C})}}_i$ with the covariant
derivatives $\Dd^{^{(\rm{C})}}_{\ad}$ and $\Dd^{^{(\rm{G})}}_{\ad}$
\beak{c}\nk\label{KDcDG}
\left[\Dd^{^{(\rm{C})}}_{\ad},\mathbf{K}^{^{(\rm{C})}}_i\right]=0~,~~\qquad\qquad
\left[\Dd^{^{(\rm{G})}}_{\ad},\mathbf{K}^{^{(\rm{C})}}_i\right]= -\,\frac{1}{2}~\th^{\a}(\s_i)_{\a\ad}~\p_t~,
\eeak
we find that Carrollian boosts preserve the C-chiral condition \eqref{Cchiral}, but not the
G-chiral condition \eqref{Gchiral}. However, notice that the deviation from G-chirality ($\th^{\a}
(\s_i)_{\a\ad}~\p_t$) is Carroll boost invariant. Therefore, we can construct two Carroll boost invariant
supersymmetric theories. The first one will have C-supersymmetry and it will be given precisely by the C-Wess-Zumino model, while the second one will have
G-supersymmetry and it will be described by the G-Wess-Zumino model with the addition of a Lagrange multiplier superfield that imposes appropriate constraints\footnote{Using \eqref{KDcDG} one can guess that the appropriate constraint should be the vanishing of the time derivative of the G-chiral superfield and that of its prepotential $\Lambda^{^{(\rm{G})}}$.} on the prepotential $\Lambda^{^{(\rm{G})}}$ of the G-chiral multiplet.

\subsection{Supersymmetric extension of electric Carroll theory}\label{seC}

By solving the C-chiral constraint \eqref{Cchiral}, we can express
$\Phi^{^{(\rm{C})}}$ in terms of an unconstrained scalar superfield, $\Lambda^{^{(\rm{C})}}$:
\beak{c}\nk
\Dd^{^{(\rm{C})}}_{\ad}\Phi^{^{(\rm{C})}}=0~~~~\rightarrow~~~~\Phi^{^{(\rm{C})}}=[\Dd^{^{(\rm{C})}}]^2\,\Lambda^{^{(\rm{C})}}~.
\eeak
Under Carrollian boosts, the unconstrained prepotential $\Lambda^{^{(\rm{C})}}$ transforms as expected:
\beak{c}\nk
\d^{^{(\rm{C})}}_{B}\Lambda^{^{(\rm{C})}}=0~~~~~\Rightarrow~~~~~\bar{\d}^{^{(\rm{C})}}_{B}\Lambda^{^{(\rm{C})}}=i\,b^i\,\mathbf{K}^{^{(\rm{C})}}_{i}\Lambda^{^{(\rm{C})}}=-\,b^i\,x_i~\p_t\,\Lambda^{^{(\rm{C})}}~,
\eeak
and because the C-covariant derivative also remains invariant \eqref{KDcDG}, the C-chiral superfield transforms as follows:
\beak{c}\nk
\d^{^{(\rm{C})}}_{B}\Phi^{^{(\rm{C})}}=0~~~~~\Rightarrow~~~~~\bar{\d}^{^{(\rm{C})}}_{B}\Phi^{^{(\rm{C})}}=i\,b^i\,\mathbf{K}^{^{(\rm{C})}}_{i}\Phi^{^{(\rm{C})}}=-\,b^i\,x_i~\p_t\,\Phi^{^{(\rm{C})}}~.
\eeak
Therefore the C-Wess-Zumino model \eqref{Cwz} is invariant under Carrollian boost and provides the C-supersymmetric extension of the electric Carroll scalar. The component action of this theory in a two component spinor description is given in \eqref{Cwz} and in four-component spinors the action takes the following form:
\begin{align}
\mathrm{S}^{^\mathrm{\,C-SUSY}}_\mathrm{\,eC}&=\frac{1}{2}\,\int dt\,d^3x\,
\Big\{  \,(\p_t\phi_R)^2\,+\,(\p_t\phi_I)^2\,+\,{F_R}^2\,+\,{F_I}^2\,-\,\bar\psi\,\g^0\p_t\,\psi \Big\}\,.
\label{super-elec action}
\end{align}
This action includes the following terms: (\emph{i}) two real electric Carroll scalar fields \eqref{electric Carroll La} $\phi_R$ and $\phi_I$ which are the real and imaginary parts of $\phi$, (\emph{ii}) two real auxiliary scalar fields $F_R$ and $F_{I}$ which are the real and imaginary parts of $F$, they have no dynamics and are required by off-shell C-supersymmetry invariance, and (\emph{iii}) an electric Carroll Majorana spinor $\psi$, which is governed by the Lagrangian\footnote{In comparison with \eqref{elec f}, here, a factor of $\frac{1}{2}$ is used due to the self-conjugacy of the Majorana field.} \eqref{elec f}. The off-shell C-supersymmetry transformations of all these fields are
\beak{ll}\nk\label{super-elec tr}
\d\phi_R=\bar\ep\,\psi \,,~\,&\d\phi_I=\bar\ep\,i\,\g^5\,\psi \,,\snk\\[8pt]
    \d F_R=-\,\bar\ep\,\g^0\,\p_t\,\psi \,, &\d F_I=-\,\bar\ep\,i\,\g^5\g^0\,\p_t\,\psi  \,,\snk\\[8pt]
    \d\psi=\g^0\p_t\,(\,\phi_R+i\,\g^5\,\phi_I\,)\,\ep - (\,F_R+i\,\g^5\,F_I\,)\,\ep \,,&\snk
\eeak
\noindent where $\ep$ is an arbitrary constant Majorana spinor that parametrize the supersymmetry transformations.
It is straightforward to check that these transformations close off-shell for every field\footnote{By
rescaling the supersymmetry transformations \eqref{super-elec tr} appropriately, we can make their supersymmetry algebra
\eqref{222} consistent with the two component convention \eqref{Csusy}.}
\be
[\,\d\1\,,\,\d\2\,]=2\,(\bar\ep\2\,\g^0\,\ep\1)\,\p_t~. \label{222}
\ee

\subsection{Supersymmetric extension of magnetic Carroll theory}

Similar to the C-chiral superfield, the G-chiral superfield can also be expressed in terms of an
unconstrained prepotential $\Lambda^{^{(\rm{G})}}$ by solving constraint \eqref{Gchiral}:
$\Phi^{^{(\rm{G})}}=[\Dd^{^{(\rm{G})}}]^2\,\Lambda^{^{(\rm{G})}}$. The prepotential $\Lambda^{^{(\rm{G})}}$ which is an unconstrained scalar superfield, transforms under Carrollian boost in the usual manner:
\beak{c}\nk
\d^{^{(\rm{C})}}_{B}\Lambda^{^{(\rm{G})}}=0~~~~~\Rightarrow~~~~~\bar{\d}^{^{(\rm{C})}}_{B}\Lambda^{^{(\rm{G})}}=-\,b^ix_i~\p_t\,\Lambda^{^{(\rm{G})}}~.
\eeak
However, as indicated by \eqref{KDcDG}, the G-covariant derivatives do transform under Carrollian boosts, therefore the transformation of $\Phi^{^{(\rm{G})}}$ is:
\beak{ll}
\d^{^{(\rm{C})}}_{B}\Phi^{^{(\rm{G})}}=-\,\frac{i}{2}~b^i\,\th^{\a}(\s_i)_{\a}{}^{\ad}~\Dd^{^{(\rm{G})}}_{\ad}\p_t\,\Lambda^{^{(\rm{G})}}~~~~~\Rightarrow~~~~~\bar{\d}^{^{(\rm{C})}}_{B}\Phi^{^{(\rm{G})}}=&\,-\,\frac{i}{2}~b^i\,\th^{\a}(\s_i)_{\a}{}^{\ad}~\Dd^{^{(\rm{G})}}_{\ad}\p_t\,\Lambda^{^{(\rm{G})}}~~~~~~~~~\nk\\[8pt]
&\,-\,b^ix_i~\p_t\,\Phi^{^{(\rm{G})}}~,
\eeak
which obviously breaks the G-chirality of $\Phi^{^{(\rm{G})}}$. Therefore, the Carroll boost invariant theory cannot be expressed purely in terms of superfield $\Phi^{^{(\rm{G})}}$ and the bare prepotential $\Lambda^{^{(\rm{G})}}$ must participate. Moreover, the Carroll boost transformation of $\Dd^{^{(\rm{G})}}_\ad\Lambda^{^{(\rm{G})}}$ is:
\beak{l}\nk
\d^{^{(\rm{C})}}_{B}\left(\Dd^{^{(\rm{G})}}_{\ad}\Lambda^{^{(\rm{G})}}\right)=-\,\frac{i}{2}~b^i\,\th^{\a}(\s_i)_{\a\ad}~\p_t\,\Lambda^{^{(\rm{G})}}~,
\eeak
hence we get the following sequence which realizes the nilpotent property of $\d^{^{(\rm{C})}}_{B}$:
\beak{c}\nk\label{GsusymC}
[\Dd^{^{(\rm{G})}}]^2\,\Lambda^{^{(\rm{G})}}~\xrightarrow{\d^{^{(\rm{C})}}_{B}} ~\p_t\,\Dd^{^{(\rm{G})}}_{\ad}\Lambda^{^{(\rm{G})}}~,~\qquad\qquad
\Dd^{^{(\rm{G})}}_{\ad}\Lambda^{^{(\rm{G})}}~\xrightarrow{\d^{^{(\rm{C})}}_{B}} ~\pa_t\,\Lambda^{^{(\rm{G})}}~\xrightarrow{\d^{^{(\rm{C})}}_{B}}~0~.
\eeak

It becomes obvious that the  deformation of the G-Wess-Zumino model under Carroll boost has a special form, it is written as the product of two factors with one of them being Carroll boost invariant. This type of deformation can be compensated by the introduction of an appropriate compensator superfield $\Sigma$. Specifically, we find that the following action is invariant under Carroll boosts and G-supersymmetry:
\beak{l}\nk\label{dBS}
\mathrm{S}^{^\mathrm{\,G-SUSY}}_\mathrm{\,mC}=\int~dt\,d^3x\,d^4\th~\left\{\,[\D^{^{(\rm{G})}}]^2\,\bar{\Lambda}^{^{(\rm{G})}}~[\Dd^{^{(\rm{G})}}]^2\,\Lambda^{^{(\rm{G})}}\,+\,\Sigma^{^{(\rm{G})}}\,\p_t\Lambda^{^{(\rm{G})}}\,+\,\bar{\Sigma}^{^{(\rm{G})}}\,\p_t\bar{\Lambda}^{^{(\rm{G})}}\right\}~,
\eeak
where $\Sigma^{^{(\rm{G})}}$ is a G-complex linear superfield $\left(\,[\Dd^{^{(\rm{G})}}]^2\,\Sigma^{^{(\rm{G})}}=0\,\right)$ and under Carrollian boosts transforms:
\beak{l}\nk
\d^{^{(\rm{C})}}_{B}\Sigma^{^{(\rm{G})}}=\frac{i}{2}~b^i\,\th^{\a}(\s_i)_{\a}{}^{\ad}~\Dd^{^{(\rm{G})}}_{\ad}[\D^{^{(\rm{G})}}]^2\,\bar{\Lambda}^{^{(\rm{G})}}~,
\eeak
which respects the complex linear nature of $\Sigma^{^{(\rm{G})}}$.
Action \eqref{GsusymC} can be written equivalently as:
\beak{l}\nk
\mathrm{S}^{^\mathrm{\,G-SUSY}}_\mathrm{\,mC}=-\int~dt\,d^3x\,d^4\th~\left\{~\Gamma^{^{(\rm{G})}}{}^{\a}\,\D^{^{(\rm{G})}}_{\a}\Dd^{^{(\rm{G})}}{}^{\ad}\,\bar{\Gamma}^{^{(\rm{G})}}_{\ad}
\,+\,\left[~\bar{\Xi}^{^{(\rm{G})}}{}^{\ad}\,\p_t\,\Gamma^{^{(\rm{G})}}_{\a}\,+\,c.c.\,\right]~\right\}~,
\eeak
where $\bar{\Gamma}^{^{(\rm{G})}}_{\ad}=\frac{1}{2}\,\Dd^{^{(\rm{G})}}_{\ad}\,\Lambda^{^{(\rm{G})}}$, and satisfies the identity $\Dd^{{^{(\rm{G})}}}_{(\bd}\bar{\Gamma}_{\ad)}=0$, while
$\bar{\Xi}^{^{(\rm{G})}}_{\ad}$ is the prepotential of $\Sigma^{^{(\rm{G})}}$, $\Sigma^{^{(\rm{G})}}=\frac{1}{2}\,\Dd^{^{(\rm{G})}}{}^{\ad}\,\bar{\Xi}^{^{(\rm{G})}}_{\ad}$, and it is defined modulo the redundancy $\Xi^{^{(\rm{G})}}_{\a}\sim\Xi^{^{(\rm{G})}}_{\a}+\D^{^{(\rm{G})}}{}^{\b}\,\tau_{\b\a}$, $\tau_{\b\a}=\tau_{\a\b}$. Using \eqref{dBS}, the transformation of $\bar{\Xi}_{\ad}$ under Carroll boosts is
\beak{l}\nk\label{dBG}
\d^{^{(\rm{C})}}_{B}\,\bar{\Xi}^{^{(\rm{G})}}_{\ad}=i~b^i\,\th^{\a}(\s_i)_{\a\ad}~\D^{^{(\rm{G})}}{}^{\b}\,\Gamma^{^{(\rm{G})}}_{\b}~.
\eeak

The corresponding component action is:
\beak{ll}\phantomsection\label{GsusyMC}
\mathrm{S}^{^\mathrm{\,G-SUSY}}_\mathrm{\,mC}=-\,\frac{1}{2}\int\,dt\,d^3x~\Big\{~&
\,(\p_i\,\pi_{R})^2\,+\,(\p_i\,\pi_{I})^2\,
-\,{G_R}^2\,-\,{G_I}^2\,+\,\bar\lambda\,\g^i\p_i\,\lambda\\[8pt]
&\,-\,2\,\chi_{1}\,\p_t\,\pi_{R}\,-\,2\,\chi_{2}\,\p_t\,\pi_{I}\,
    +\,B_1\,\p_t\,G_{R}\,+\,B_2\,\p_t\,G_{I}\,+\,\bar{\rho}\,\g^0\p_t\,\lambda
    +\bar{\lambda}\,\g^0\p_t\,\rho\,\\[8pt]
&\,+\,\bar{\beta}\,\g^0\p_t\,\Upsilon\,+\bar{\Upsilon}\,\g^0\p_t\,\beta
\,+\,U_1\,\p_t\,\Lambda^{R}+\,U_2\,\p_t\,\Lambda^{I}\,\\[8pt]
&\,+\,U_1^{i}\,\p_t\,\Lambda^{R}_{i}+\,U_2^{i}\,\p_t\,\Lambda^{I}_{i}\,
+\,\bar{\eta}\,\g^0\p_t\,\Omega\,+\bar{\Omega}\,\g^0\p_t\,\eta
~\Big\}~.\nk
\eeak
The terms in the first line are the G-Wess-Zumino terms \eqref{Gwz} consisting of: (\emph{i}) the magnetic Galilei Lagrangian \eqref{magnetic Galil} for two real scalar fields  $\pi_R$ and $\pi_I$ which are the real and imaginary parts of $\pi$, (\emph{ii}) two real auxiliary scalar fields, $G_R$ and $G_I$, the real and imaginary parts of $G$, which have no dynamics but are required by G-supersymmetry, and (\emph{iii}) a magnetic Galilei Lagrangian \eqref{elec f2} for Majorana field $\lambda$. The terms in the second line, are the Lagrange multiplier terms that impose appropriate constraints and make the theory invariant under Carroll boosts. The first two terms are the constraint terms for the two scalars $\pi_R,\,\pi_I$, as expected by \eqref{mcL}, the following two terms are the analog terms for the auxiliary fields $G_R,\,G_I$, and the last two terms are the constraint terms for the fermion $\lambda$ as found in \eqref{magn f}.
Finally, the terms in the third and forth lines are required by the supersymmetry transformation of the terms in the second line. Specifically, the component fields that appear in the above action are organized in the following supermultiplets $\Gamma^{^{(\rm{G})}}$ and $\Xi^{^{(\rm{G})}}$ in the following manner:
\beak{ll}\nk
\Gamma^{^{(\rm{G})}}:=\Big\{\Lambda_{R},\,\Lambda_{I},\,\Lambda^{i}_{R},\,\Lambda^{i}_{I},\,\pi_{R},\,\pi_{I},\,G_{R},\,G_{I}\,&\mid\,\Upsilon,\,\lambda,\,\Omega\Big\}~,\\[8pt]
\Xi^{^{(\rm{G})}}:=\Big\{\underbrace{\addstackgap[6pt]{B_{1},\,B_{2},\,U_{1},\,U_{2},\,U^{i}_{1},\,U^{i}_{2},\,\chi_1,\,\chi_2\,}}_{\text{Bosons}}&\mid\underbrace{\addstackgap[6pt]{\,\,\rho,\,\eta,\,\beta}}_{\text{Fermions}}\Big\}~.
\eeak
Each multiplet carries 12 bosonic and 12 fermionic off-shell degrees of freedom. The supersymmetry transformations of these fields are easily derived from their corresponding supermultiplets and they generalise the transformations \eqref{smGtr} of the smaller G-Chiral multiplet $\{\pi_R,\,\pi_I,\,G_R,\,G_I,\,\lambda\}$. The closure of these transformations is consistent with G-supersymmetry:
\be
[\,\d\1\,,\,\d\2\,]=2\,(\bar\ep\2\,\g^i\,\ep\1)\,\p_i~.\label{gsusy}
\ee


\section{Galilei Boost Invariant Supersymmetric Theories} \label{sec6}
Now, we study the transformation properties of the C-Wess-Zumino and G-Wess-Zumino models under Galilean
boosts, generated by $\mathbf{K}^{^{(\rm{G})}}_i$. The commutator of $\mathbf{K}^{^{(\rm{G})}}_i$ with the supersymmetric covariant
derivatives $\Dd^{^{(\rm{C})}}_{\ad}$ and $\Dd^{^{(\rm{G})}}_{\ad}$:
\beak{c}\nk\label{KGDcDG}
\left[\Dd^{^{(\rm{C})}}_{\ad},\mathbf{K}^{^{(\rm{G})}}_i\right]=-\,\frac{1}{2}~\th^{\a}(\s^0)_{\a\ad}~\p_i~,~~\qquad\qquad
\left[\Dd^{^{(\rm{G})}}_{\ad},\mathbf{K}^{^{(\rm{G})}}_i\right]= 0~,
\eeak
signals that the Galilean boosts preserve the G-chiral condition \eqref{Gchiral}, but not C-chirality \eqref{Cchiral}. Hence, the supersymmetric extension of the magnetic Galilei scalar theory will be provided by the G-Wess-Zumino model \eqref{Gwz} which is manifestly invariant under G-supersymmetry, whereas the supersymmetric extension of the electric Galilei scalar theory will be given by the C-Wess-Zumino model \eqref{Cwz} which is manifestly C-supersymmetric.
\subsection{Supersymmetric extension of magnetic Galilei theory}
Repeating the same type of arguments as in section \ref{seC} and using \eqref{KGDcDG} we conclude that
the G-chiral superfield $\Phi^{^{(\rm{G})}}$ is invariant under Galilean boosts:
\beak{c}\nk
\d^{^{(\rm{G})}}_{B}\Phi^{^{(\rm{G})}}=0~~~~~~\Rightarrow~~~~~~\bar{\d}^{^{(\rm{G})}}_{B}\Phi^{^{(\rm{G})}}=i\,b^i\,\mathbf{K}^{^{(\rm{G})}}_{i}\Phi^{^{(\rm{G})}}=-\,b^i\,t\,\p_i\,\Phi^{^{(\rm{G})}}~.
\eeak
Hence, the G-Wess-Zumino model is invariant under Galilean boost as well as G-supersymmetry:
\beak{l}\nk
\mathrm{S}^{^\mathrm{\,\rm{G}-SUSY}}_\mathrm{\,mG}\sim \int dt\,d^3x\,d^4\th~\bar{\Phi}^{^{(\rm{G})}}\,\Phi^{^{(\rm{G})}}~.
\eeak
The component action of this theory in four component description is
\be
\mathrm{S}^{^\mathrm{\,\rm{G}-SUSY}}_\mathrm{\,mG}=-\,\frac{1}{2}\,\int dt\,d^3x\,
\Big\{  \,(\p_i\pi_{R})^2\,+\,(\p_i\pi_{I})^2\,-\,{G_R}^2\,-\,{G_I}^2\,+\,\bar\lambda\,\g^i\p_i\,\lambda \Big\}\,.
\label{super-magne action}
\ee
The action includes the following terms: (\emph{i}) the Lagrangians of two real magnetic Galilei scalar fields \eqref{magnetic Galil} $\phi_{R}$ and $\phi_{I}$, (\emph{ii}) two real auxiliary scalar fields $G_R$ and $G_I$ which have no dynamics and are required by off-shell
G-supersymmetry, and (\emph{iii}) a magnetic Galilei Majorana field $\lambda$, which is described by the Lagrangian \eqref{elec f2}. The explicit G-supersymmetry transformations of all the fields are
\beak{ll}\nk\label{smGtr}
     \d\pi_{R}=\bar\ep\,\lambda \,,~\,&\d\pi_I=\bar\ep\,i\,\g^5\,\lambda \,,\snk\\[8pt]
    \d G_R=\bar\ep\,\g^i\,\p_i\,\lambda \,, &\d G_I=\bar\ep\,i\,\g^5\g^i\,\p_i\,\lambda  \,,\snk\\[8pt]
    \d\lambda=\g^i\p_i\,(\,\pi_{R}+i\,\g^5\,\pi_{I}\,)\,\ep + (\,G_{R}+i\,\g^5\,G_I\,)\,\ep \,.\snk
\eeak
where $\ep$ is an arbitrary constant Majorana spinor object that parametrizes the supersymmetry transformations. It is straightforward to check that the algebra of the above transformations indeed closes off-shell to the expected G-supersymmetry, for all fields:
\be
[\,\d\1\,,\,\d\2\,]\,=2\,(\bar\ep\2\,\g^i\,\ep\1)\,\p_i\,.
\ee
Systems that respect this type of supersymmetry have been studied in \cite{Gates:2013caa}.
\subsection{Supersymmetric extension of electric Galilei theory}
Using the seed Lagrangian approach \cite{Bergshoeff:2022qkx}, it was shown that an electric description of Galilean scalar can be obtained (see \ref{eGs} for review) by considering the electric Carrollian description and adding an appropriate Lagrange multiplier term that enforces Galilei boost invariance and imposes the appropriate constraints.
The supersymmetric extension of that theory is straightforward. The electric Carrollian scalar acquires a C-supersymmetric partner, the electric Carroll fermion \eqref{elec f}. Together with the appropriate auxiliary fields they make the C-Wess-Zumino model \eqref{Cwz} described by a C-chiral superfield $\Phi^{^{(\rm{C})}}$. Similarly, the bosonic and fermionic Lagrange multipliers that appear in the electric description of the Galilean bosons \eqref{eGL} and fermions \eqref{eG} respectively, will be organized into a second C-supermultiplet.

\vspace{.3cm}

Using \eqref{KGDcDG}, we calculate the transformation of the prepotential $\Lambda^{^{(\rm{C})}}$ and its derivatives $\Dd^{^{(\rm{C})}}_{\ad}\Lambda^{^{(\rm{C})}}$,\, $\Phi^{^{(\rm{C})}}=[\Dd^{^{(\rm{C})}}]^2\,\Lambda^{^{(\rm{C})}}$ under Galilei boosts:
\beak{ll}\nk
\d^{^{(\rm{G})}}_{B}\Lambda^{^{(\rm{C})}}=0~~\Rightarrow~~\bar{\d}^{^{(\rm{G})}}_{B}\Lambda^{^{(\rm{C})}}=-\,b^i\,t\,\p_i\,\Lambda^{^{(\rm{C})}}~,&\snk\\[8pt]
\d^{^{(\rm{G})}}_{B}\left(\Dd^{^{(\rm{C})}}_{\ad}\Lambda^{^{(\rm{C})}}\right)=-\,\frac{i}{2}~b^i\,\th^{\a}(\s^0)_{\a\ad}\,\p_i\,\Lambda^{^{(\rm{C})}}~,&\snk\\[8pt]
\d^{^{(\rm{G})}}_{B}\Phi^{^{(\rm{C})}}=-\,\frac{i}{2}\,b^i\,\th^{\a}\,(\s^0)_{\a}{}^{\ad}~\Dd^{^{(\rm{C})}}_{\ad}\p_i\,\Lambda^{^{(\rm{C})}}~~~~~\Rightarrow~~~~~\bar{\d}^{^{(\rm{G})}}_{B}\Phi^{^{(\rm{C})}}=&\,-\,\frac{i}{2}\,b^i\,\th^{\a}(\s^0)_{\a}{}^{\ad}\,\Dd^{^{(\rm{C})}}_{\ad}\p_i\,\Lambda^{^{(\rm{C})}}~~~~\snk\\[8pt]
&\,-\,b^i\,t\,\p_i\,\Phi^{^{(\rm{C})}}~,
\eeak
These transformations realize the nilpotency of the Galilean boost transformation $\d^{^{(\rm{G})}}_{B}$
in the sequence
\beak{c}\nk\label{CsusyChain}
[\Dd^{^{(\rm{C})}}]^2\,\Lambda^{^{(\rm{C})}}~\xrightarrow{\d^{^{(\rm{G})}}_{B}} ~\p_i\,\Dd^{^{(\rm{C})}}_{\ad}\,\Lambda^{^{(\rm{C})}}~,~\qquad\qquad
\Dd^{^{(\rm{C})}}_{\ad}\Lambda^{^{(\rm{C})}}~\xrightarrow{\d^{^{(\rm{G})}}_{B}} ~\pa_i\,\Lambda^{^{(\rm{C})}}~\xrightarrow{\d^{^{(\rm{G})}}_{B}}~0~.
\eeak

Applying the seed Lagrangian approach to superspace, we start with the C-Wess-Zumino model
$\mathrm{S}^{^\mathrm{\,\rm{C}-SUSY}}_\mathrm{\,eC}\sim\int dt\,d^3x\,d^4\th~\bar{\Phi}^{^{(\rm{C})}}\,\Phi^{^{(\rm{C})}}$ and transform it under Galilean boosts:
\beak{l}\nk
\d^{^{(\rm{G})}}_{B}\mathrm{S}^{^\mathrm{\,\rm{C}-SUSY}}_\mathrm{\,eC}=\,-\,
\int dt\,d^3x\,d^4\th~\Big\{\, \Dd^{^{(\rm{C})}}{}^{\ad}\,\Phi^{^{(\rm{C})}}~\frac{i}{2}\,b^i\,\th^{\a}(\s^0)_{\a\ad}~\p_i\,\Lambda^{^{(\rm{C})}}\,\Big\}~.
\eeak
Once again the deformation of the theory is factorized and one of the factors ($\,\p_i\,\Lambda^{^{(\rm{C})}}$\,) is G-boost invariant, as seen in \eqref{CsusyChain}. Therefore, the deformation can be absorbed by introducing a collection of compensator superfields $\Sigma^{^{(\rm{C})}}_{i}$ with appropriately chosen transformations under Galilean boosts. The following action is invariant under Galilei boosts and C-supersymmetry:
\beak{l}\nk\label{CsusyeG}
\mathrm{S}^{^\mathrm{\,C-SUSY}}_\mathrm{\,eG}=\int~dt\,d^3x\,d^4\th~\left\{[\D^{^{(\rm{C})}}]^2\,\bar{\Lambda}^{^{(\rm{C})}}~[\Dd^{^{(\rm{C})}}]^2\,\Lambda^{^{(\rm{C})}}\,+\,\Sigma^{^{(\rm{C})}}{}^{i}\,\p_i\,\Lambda^{^{(\rm{C})}}\,+\,\bar{\Sigma}^{^{(\rm{C})}}{}^{i}\,\p_i\,\bar{\Lambda}^{^{(\rm{C})}}\right\}~,~~~~
\eeak
where $\Sigma^{^{(\rm{C})}}_{i}$ are C-complex linear superfield $\left(\,[\Dd^{^{(\rm{C})}}]^2\,\Sigma^{^{(\rm{C})}}_{i}=0\,\right)$ and under Galilei boosts transform as follows:
\beak{l}\nk
\d^{^{(\rm{G})}}_{B}\Sigma^{^{(\rm{C})}}_i=\frac{i}{2}~b_i\,\th^{\a}(\s^0)_{\a}{}^{\ad}~\Dd^{^{(\rm{C})}}_{\ad}\,[\D^{^{(\rm{C})}}]^2\,\bar{\Lambda}^{^{(\rm{C})}}~.
\eeak
The action \eqref{CsusyeG} can be expressed in terms of the variables $\bar{\Gamma}^{^{(\rm{C})}}_{\ad}$ and $\bar{\Xi}^{^{(\rm{C})}}_i{}_{\ad}$ defined as:
\beak{l}\nk
\bar{\Gamma}^{^{(\rm{C})}}_{\ad}=\frac{1}{2}\,\Dd^{^{(\rm{C})}}_{\ad}\,\Lambda^{^{(\rm{C})}}~,~\qquad\qquad
\Sigma^{^{(\rm{C})}}_i=\frac{1}{2}\,\Dd^{^{(\rm{C})}}{}^{\ad}\,\bar{\Xi}^{^{(\rm{C})}}{}_i{}_{\ad}~,
\eeak
where $\bar{\Xi}^{^{(\rm{C})}}{}_i{}_{\ad}$ is defined up modulo the redundancy $\Xi^{^{(\rm{C})}}_i{}_{\a}\sim\Xi^{^{(\rm{C})}}_i{}_{\a}+\D^{^{(\rm{C})}}\tau_{\b\a},\, \tau_{\a\b}=\tau_{\b\a}$,
\beak{l}\nk
\mathrm{S}^{^\mathrm{\,C-SUSY}}_\mathrm{\,eG}=-\int~dt\,d^3x\,d^4\th~\left\{~\Gamma^{^{(\rm{C})}}{}^{\a}\,\D^{^{(\rm{C})}}_{\a}\Dd^{^{(\rm{C})}}{}^{\ad}\,\bar{\Gamma}^{^{(\rm{C})}}_{\ad}
\,+\,\left[~\bar{\Xi}^{^{(\rm{C})}}{}^{i\,\ad}\,\p_i\,\Gamma^{^{(\rm{C})}}_{\a}+c.c.\,\right]~\right\}~.
\eeak

The component action of this theory can be extracted from the above superspace action:
\beak{ll}\phantomsection\label{CsusyEG}
\mathrm{S}^{^\mathrm{\,C-SUSY}}_\mathrm{\,eG}=\,\frac{1}{2}\int\,dt\,d^3x~\Big\{&
\,(\p_t\,\phi_{R})^2\,+\,(\p_t\,\phi_{I})^2\,+\,{F_R}^2\,+\,{F_I}^2\,-\,\bar\psi\,\g^0\p_t\,\psi\\[8pt]
&\,-\,2\,\chi^i_{1}\,\p_i\,\phi_{R}\,-\,2\,\chi^i_{2}\,\p_i\,\phi_{I}\,
+\,B^i_1\,\p_i\,F_{R}\,+\,B^i_2\,\p_i\,F_{I}\,-\,\bar{\rho}^{\,i}\,\g^0\,\p_i\,\psi\,
-\,\bar{\psi}\,\g^0\,\p_i\,\rho^i\\[8pt]
&\,+\,\,\bar{\beta}^{\,i}\,\g^0\,\p_i\,\Upsilon\,+\bar{\Upsilon}\,\g^0\,\p_i\,\beta^{\,i}
\,+\,U_1^{i}\,\p_i\,\Lambda_{R}+\,U_2^{i}\,\p_i\,\Lambda_{I}\\[8pt]
&\,+\,\,(U_1)^{i}{}_j\,\p_i\,\Lambda_{R}^j+\,(U_2)^{i}{}_j\,\p_i\,\Lambda_{I}^j\,+\,\bar{\eta}^{\,i}\,\g^0\,\p_i\,\Omega\,+\bar{\Omega}\,\g^0\,\p_i\,\eta^i
\Big\}~.\nk
\eeak
The terms on the first line are the C-Wess-Zumino terms \eqref{Cwz} consisting of: (\emph{i}) the electric Carroll Lagrangians \eqref{electric Carroll La} for two real electric Carroll
scalar fields  $\phi_R$ and $\phi_I$ which are the real and imaginary parts of $\phi$, (\emph{ii}) two real auxiliary scalar fields, $F_R$ and $F_I$, the real and imaginary parts of $F$, which have no dynamics but are required by C-supersymmetry, and (\emph{iii}) an electric Carroll Lagrangian \eqref{elec f} for Majorana field $\psi$. The terms on the second line,
are the Lagrange multiplier terms that impose appropriate constraints and make the theory invariant under Galilei boosts. The first two terms are the constraint terms for the two scalars $\phi_R,\,\phi_I$, as expected by \cite{Bergshoeff:2022qkx} (see \eqref{eGL}), the following
two terms are the analog terms for the auxiliary fields $F_R,\,F_I$, and the last two terms are the constraint terms for the fermion $\psi$ as found in \eqref{eG}.
Finally, the third and fourth line terms are required by the supersymmetry transformations of the second line terms. The component fields are organized into four supermultiplets $\Gamma^{^{(\rm{C})}}$ and $\Xi^{^{(\rm{C})}}{}^i$, $i=1,2,3$ as follows:
\beak{l}\nk
~\Gamma^{^{(\rm{C})}}:=\Big\{\Lambda_{R},\,\Lambda_{I},\,\Lambda_{R}^{i},\,\Lambda_{I}^{i},\,\phi_{R},\,\phi_{I},\,F_{R},\,F_{I}\,\mid\,\Upsilon,\,\psi,\,\Omega\Big\}\\[8pt]
\Xi^{^{(\rm{C})}}{}^i:=\Big\{\underbrace{\addstackgap[6pt]{B^i_{1},\,B^i_{2},\,U^i_{1},\,U^i_{2},\,(U_{1})^i{}_j,\,(U_{2})^i{}_j,\,\chi^i_1,\,\chi^i_2\,}}_{\text{Bosons}}\mid\underbrace{\addstackgap[6pt]{\,\rho^i,\,\eta^i,\,\beta^i}}_{\text{Fermions}}\Big\}~.
\eeak
Each one of these supermultiplets has 12 bosonic and 12 fermionic d.o.f. and one can easily check that their supersymmetry transformations close off-shell to the C-supersymmetry algebra
\be
[\,\d\1\,,\,\d\2\,]=2\,(\bar\ep\2\,\g^0\,\ep\1)\,\p_t~.
\ee
As mentioned previously supersymmetric Carrollian multiplets can be understood as supersymmetric quantum mechanical systems. In this case, the detailed analysis of the quantum mechanical complex linear supermultiplet, which corresponds to the Carrollian multiplet $\Sigma^{^{(\rm{C})}}$ can be found in
\cite{Gates:2012zr,Gates:2011aa,Caldwell:2017ejk}.

\section{Conclusions and outlook}\label{conclu}

In this paper, we have studied the non-Lorentzian algebras and theories associated with the Carroll and Galilei symmetries, and we have addressed the following important issues.

\vspace{.3cm}

$\bullet$ We explored the construction of super-Carroll and super-Galilei algebras
through the introduction of contraction parameters in the $4D, \mathcal{N}=1$ super-Poincar\'{e} algebra. These algebras exhibit a breaking of Lorentz symmetry due to a characteristic distinction between temporal and spatial components which is generated by a difference in the associated exponents of the contraction parameter. To ensure the proper behavior of these algebras in the limits of $c\to 0$ (Carroll) and $c\to \infty$ (Galilei), we imposed specific conditions on these exponents. These conditions, ultimately give rise to the super-Carroll and super-Galilei algebras, which have been described in detail, including their generators \eqref{scgenerators}, \eqref{sggenerators} and commutation relations \eqref{sCa}, \eqref{sGa} respectively.

\vspace{.3cm}

$\bullet$ Furthermore, we discussed the action of Carrollian and Galilean supersymmetries, as well as Carrollian and Galilean boosts on coordinates and (super)fields. Notably, both Carrollian and Galilean boost transformations of coordinates were shown to be nilpotent. To facilitate the formulation of theories with manifest Carrollian or Galilean supersymmetry, we introduced the corresponding superspace covariant derivatives \eqref{ccderivative}, \eqref{gcderivative}. These derivatives play an important role in understanding the behavior of the theory under Carrollian or Galilean boosts within a supersymmetric framework.

\vspace{.3cm}

$\bullet$ From the field theoretical point of view, the scalar field theories with Carroll and Galilei symmetries have been investigated, and they exhibit electric and magnetic sectors \cite{Henneaux:2021yzg, deBoer:2021jej, Bergshoeff:2022qkx}. The electric sector corresponds to the theory where the time derivatives are dominant, while the magnetic sector is the theory in which the spatial derivatives are dominant. We reviewed four existing methods to construct the Carroll scalar field theories in appendix \ref{Carroll scalar rev}: the Hamiltonian formulation, the field expansion approach, the Lagrange multiplier technique, and the seed Lagrangian method. For the Galilei scalar field theory, only the seed Lagrangian method has been used in the literature \cite{Bergshoeff:2022qkx}, but we extend it to the other three methods in appendix \ref{Gscalar}.

\vspace{.3cm}

$\bullet$ Motivated by the success of the four methods employed in the scalar field theory, our investigation naturally expanded to encompass the construction of Carroll and Galilei fermionic field theories, including Dirac, Majorana, and Weyl fermions. We used three methods, except the Lagrange multiplier method which is not suitable for fermions due to the nonquadratic kinetic term, to construct the Carroll and Galilei fermionic field theories in sections \ref{Cfermions} and \ref{Gfermions} respectively. Two important points that facilitate these constructions are:  (\emph{i}) Carroll and Galilei fermions transform like a scalar under Carroll/Galilei boosts as a result of the algebra contraction process, and (\emph{ii})
rather than incorporating the contraction parameter within the metric as in \cite{Banerjee:2022ocj,Bagchi:2022eui}\footnote{The authors of \cite{Banerjee:2022ocj, Bagchi:2022eui} incorporate $c$ into the metric, which requires them to introduce new gamma matrices and Clifford algebra. As a result, their Carroll fermionic field theories are different from ours.}, we opted to include it within the component coordinates; thus, we kept the Minkowski metric, the Clifford algebra, and gamma matrices unchanged. While this consideration has no significant effect in the scalar field theory, it has an important effect in the fermionic theories and allows us to use the same gamma matrices as in relativistic theory and consequently led us to consistent fermionic theories with supersymmetry.

\vspace{.3cm}

$\bullet$ Moreover, we investigated the supersymmetric extension of these models.
There are two types of non-Lorentzian supersymmetry that one can consider, \emph{C-supersymmetry} \eqref{Csusy} and \emph{G-supersymmetry} \eqref{Gsusy}.
We found that a natural description of simple matter theories that are manifestly invariant under C or G supersymmetry, is provided  by the C-Wess-Zumino \eqref{Cwz} and G-Wess-Zumino \eqref{Gwz} models. These models combine the known descriptions of
Carroll/Galilei scalars with the newly found Carroll/Galilei fermions and promote them into appropriate supermultiplets with the addition of appropriate auxiliary fields.
The superspace description of these theories is provided by the definition of appropriate
C-chiral superfields \eqref{Cchiral} and G-chiral superfields \eqref{Gchiral}. In addition, we provided the component description of these models and the detailed supersymmetry transformation laws for all the fields.

\vspace{.3cm}

$\bullet$ The structure of the component description of C-Wess-Zumino model makes obvious that it corresponds to the C-supersymmetrization of \emph{electric} Carrollian scalar and fermionic fields. On the other hand, the G-Wess-Zumino model provides the G-supersymmetrization of \emph{magnetic} Galilean scalars and fermions. Naturally, we searched for the supersymmetric extensions of the magnetic Carrollian and electric Galilean theories. By applying the seed Lagrangian methodology, presented in \cite{Bergshoeff:2022qkx}, in superspace and using G-Wess-Zumino and C-Wess-Zumino theories as seeds we found the corresponding magnetic Carroll \eqref{GsusyMC} and electric Galilei \eqref{CsusyEG} theories. It is interesting to emphasize that the super magnetic Carroll theory is invariant under G-supersymmetry, whereas the super electric Galilei theory is invariant under C-supersymmetry. However, it is important to emphasize that in both cases there can be supersymmetry breaking solutions where the supersymmetric auxiliary fields acquire a nontrivial vev.

\vspace{.3cm}

In the future, we would like to investigate the C/G supersymmetric extensions of non-Lorentzian gauge theories. A lot of work has be done for non-Lorentzian (super)gravities \cite{Henneaux:1979vn,Andringa:2013mma,Gallegos:2019icg,Bergshoeff:2021bmc,Bergshoeff:2021tfn} (for a review see \cite{Bergshoeff:2022iyb} and references therein) but not much is known about non-Lorentzian higher spin theories\footnote{There is a discussion of spin $s=3$ theories in \cite{Henneaux:2021yzg}.} and their C/G supersymmetric generalizations. Another interesting direction is the exploration of \emph{variant representations} of non-Lorentzian supersymmetries. The existence of variant supersymmetric representations is a characteristic feature of supersymmetric theories in various dimensions and it is often related to special features of target space geometry or the existence of (weak/strong) dualities. Therefore, it would be interesting to search for variant descriptions of non-Lorentzian matter or gauge theories.
For the case of C-supersymmetry such a search can be assisted by the \emph{Adinkras} \cite{Faux:2004wb,Doran:2006it,Gates:2009me,Zhang:2011np} methodology of classifying one dimensional representations of supersymmetry algebra. Specifically, by interpreting C-supersymmetry \eqref{Csusy} not as the outcome of a Lorentz breaking contraction of the Super-Poincar\'{e} algebra, but as its one dimensional (QM) reduction, one can use \emph{Adinkras} to discover such variant representations of C-supersymmetry and extended C-supersymmetries. Another interesting direction is the
Carroll/Fracton correspondence, which has been studied in recent works (see, e.g.,
\cite{Figueroa-OFarrill:2023qty, Figueroa-OFarrill:2023vbj} and references therein). It would be interesting
to investigate the existence of a similar type correspondence for Galilean theories.

	\section*{Acknowledgments}

The authors thank H. R. Afshar, S. J. Gates Jr, M. Khorrami and J-H. Park for helpful discussions and insightful comments. Also, the authors would like to thank M.~Henneaux for early comments. The work of K.K. is supported in part by the endowment of the Clark Leadership Chair in Science at the University of Maryland, College Park. K.K. gratefully acknowledge the hospitality of the Physics Department at the University of Maryland, College Park. The work of M.N. is partially supported by IPM funds. He also thanks the Iran National Science Foundation (INSF) for supporting this research project (Grant Number: 4000132).

	\appendix

	\section
	{Conventions}\label{conven}

We use the mostly plus signature for the Minkowski metric $\eta _{\mu \nu }={\text{diag}}(-1,1,1,1)$. Therefore, when relations are expressed with the speed of light $c$ involved, we include it in the coordinate component, i.e. $x^\m:=(ct, x^i)$, and keep the Minkowski metric $\eta_{\m\n}$, and consequently the Clifford algebra $\{\,\g^\m\,,\,\g^\n\,\}=2\e^{\m\n}$ unchanged. The following properties hold for the gamma matrices and the fifth gamma $\gamma^5 ( = i\gamma^0\gamma^1\gamma^2\gamma^3 )$:
 \be
 (\g^0)^\dagger=-\,\g^0\,,\qquad (\g^i)^\dagger=\g^i\,,\qquad
 (\g^5)^\dagger=\g^5\,,\qquad(\g^0)^2=-\,1\,\,,\qquad (\g^i)^2=1\,,\qquad (\g^5)^2=1\,.  \nonumber
 \ee

\section{Carroll scalar field} \label{Carroll scalar rev}
In this appendix, we review the Carroll scalar field theories derived through different methods, such as the Hamiltonian formulation \cite{Henneaux:2021yzg}, the field expansion and the Lagrange multiplier approaches \cite{deBoer:2021jej}, and the seed Lagrangian method \cite{Bergshoeff:2022qkx}. It is worth noting that while these methods differ in their approaches, they are fundamentally equivalent to each other.

\subsection{Hamiltonian method}	\label{hf}

The Hamiltonian formulation consists of two independent variables (the field $\phi$ and the canonical momentum conjugate to the field $\pi_\phi$), which allows us to take the Carroll limit in two different ways: explicit limit and the limit after rescaling the fields \cite{Henneaux:2021yzg}. To this end, let us consider a massless scalar field in Minkowski spacetime with the Lagrangian density
\be
\mathcal{L}=-\,\frac{1}{2}\,\p_\m\phi\,\p^\m\phi=\frac{1}{2c^2}\,(\p_t\phi)^2-\frac{1}{2}\,(\p_i\phi)^2\,. \label{Scalar Lagrangian}
\ee
The canonical momentum conjugate to the field $\phi$ reads
\be
\pi_\phi=\frac{\p\mathcal{L}}{\p\dot{\phi}}=\frac{1}{c^2}\,\dot{\phi}\,, \qquad\quad \dot{\phi}\equiv\p_t\phi\,,
\label{conjugate b}
\ee
and then the Hamiltonian density becomes
\be
\mathcal{H}=\pi_\phi\,\dot{\phi}-\mathcal{L}=\frac{1}{2}\,\Big[\,c^2(\pi_\phi)^2+(\p_i\phi)^2 \,\Big]\,.\label{Hamiltonian b}
\ee
Accordingly, the Hamiltonian action of the massless scalar field can be written as
\be
S\,[\phi,\pi_\phi]=\int\,dt\,d^3x\,\Big\{\,\pi_\phi\,\dot{\phi}- \mathcal{H} \,\Big\}=\int\,dt\,d^3x\,\Big\{ \,\pi_\phi\,\dot{\phi}-\,\frac{1}{2}\,c^2(\pi_\phi)^2\,-\,\frac{1}{2}\,(\p_i\phi)^2\,\Big\}\,.
\label{scalar Hami}
\ee
There are two ways to take the Carroll limit $c\rightarrow0$. First, we may directly take the limit of the visible $c$'s in \eqref{scalar Hami}, without rescaling the fields. However, this procedure violates the canonical relation between the field and its conjugate momentum, as seen from \eqref{conjugate b}. Therefore, after taking the limit, one should replace $\pi_\phi$ by an auxiliary field $\pi_\varphi$ in \eqref{scalar Hami} to reflect the breakdown of the canonical relation. This yields the magnetic Carroll scalar action in the Hamiltonian formulation
\be
S_\mathrm{mC}=\int\,dt\,d^3x\,\Big\{ \pi_\varphi\,\dot{\phi}-\frac{1}{2}\,(\p_i\phi)^2\,\Big\}\,.
\label{scalar magnetic}
\ee
Alternatively, we can rescale fields in \eqref{scalar Hami}, i.e. $\phi\rightarrow c\, \phi'$ and $\pi_\phi\rightarrow \tfrac{1}{c}\,\pi_\phi'$, which preserves the canonical structure. Particularly the canonical relation \eqref{conjugate b} becomes $\pi_\phi'=\dot{\phi}'$. Therefore, in this way, we first perform the rescaling, take the limit $c\rightarrow0$, and remove the primes from the rescaled variables. This leads to the electric Carroll scalar action in the Hamiltonian formulation
\be
S_\mathrm{eC}=\int\,dt\,d^3x\,\le\{ \pi_\phi\,\dot{\phi}-\frac{1}{2}\,(\pi_\phi)^2\,\ri\}\,.
\ee

\subsection{Field expansion method}\label{fe}

The field expansion technique \cite{deBoer:2021jej} is another approach to derive Carroll scalar field Lagrangians. The Lagrangian density of the massless scalar field, with the speed of light $c$ involved, is given by
\begin{align}
\mathscr{L}&=-\,\frac{1}{2}\,\p^\m{\varPhi}\,\p_\m{\varPhi}=\frac{1}{2c^2}\,(\p_t{\varPhi})^2-\frac{1}{2}\,(\p_i{\varPhi})^2\,.\label{scalar}
\end{align}
Let us now make an expansion around $c=0$ in the scalar field\footnote{Here, only even powers of $c$ contribute to terms that are Carroll boost invariant.}
\be
{\varPhi}=c^\a\Big(\phi+c^2\,\varphi+\mathcal{O}(c^4)\Big)\,,\label{Phi expansion}
\ee
for some $\a$, and define $\mathcal{L}_\mathrm{eC}$ and $\mathcal{L}_\mathrm{mC}$ trough the Lagrangian density
\be
\mathscr{L}=c^{2\a-2}\Big( \mathcal{L}_\mathrm{eC}+c^2\,\mathcal{L}_\mathrm{mC}+\mathcal{O}(c^4)\Big)\,.\label{Lagrange expan}
\ee
By applying \eqref{Phi expansion} to \eqref{scalar} and comparing with \eqref{Lagrange expan}, one can derive the electric and magnetic Carroll scalar Lagrangian densities, respectively
\begin{align}
     \mathcal{L}_\mathrm{eC}&=\frac{1}{2}\,(\p_t\phi)^2\,,\label{elec}\\[8pt]
      \mathcal{L}_\mathrm{mC}&=(\p_t\varphi)(\p_t\phi)-\,\frac{1}{2}(\p_i\phi)^2\,.\label{magn}
\end{align}

Let us discuss the invariance of the electric and magnetic Lagrangians under a Carroll boost, which is an ultrarelativistic limit of a Lorentz boost. Under a Lorentz boost, with $\vec{\b}$ being the Lorentz boost parameter, the field $\varPhi$ in \eqref{scalar} transforms as
\be
\d\,\varPhi=i\,\b_j\,J^{0j}\,\varPhi=\b_j\le(x^0\p^j-x^j\p^0\ri)\varPhi
=\le(c\,t\,\vec{\b}\c\vec{\p}+\frac{1}{c}\,\vec{\b}\c \vec{x}\, \p_t\ri)\varPhi\,, \label{Lorentz tr}
\ee
and the Lagrangian density \eqref{scalar} transforms into a total derivative. Then, one may use the scalar field expansion \eqref{Phi expansion} for \eqref{Lorentz tr} and define $\vec{\b}=c\,\vec{b}$, with $\vec{b}$ being the Carroll boost parameter. This yields the Carroll boost transformations
\begin{align}
    \d^{^{(\rm{C})}}_{B}\phi={b}^i\,{x}_i\,\p_t\,\phi\,,\qquad\qquad \d^{^{(\rm{C})}}_{B}\varphi={b}^i\,{x}_i\,\p_t\,\varphi+t\,{b}^i\,{\p}_i\,\phi\,, \label{Carroll trans}
\end{align}
under which the electric \eqref{elec} and magnetic \eqref{magn} Carroll scalar Lagrangians are invariant up to total derivatives. 

\vspace{.3cm}

The equations of motion for the electric \eqref{elec} and magnetic \eqref{magn} Lagrangians read as follows:
\begin{align}
 \mathrm{eC}-e.o.m:& \qquad  \p_t^2\,\phi=0\,, \label{eom e}\\[4pt]
 \mathrm{mC}-e.o.m:& \qquad  \p_t^2\,\phi=0\,, \qquad \p_t^2\,\varphi-\p_i^2\,\phi=0\,. \label{eom m}
\end{align}
It is useful to express two magnetic scalar field equations \eqref{eom m} in terms of a matrix equation
\be
 \qquad \mathrm{B}\,\mathrm{\Phi}=0\,, \quad\qquad \mbox{with}\quad \mathrm{B}:=\begin{pmatrix} -\,\p_i^2 & \p_t^2 \\ ~~\,\p_t^2 & 0  \end{pmatrix}\,,\quad \mathrm{\Phi}:=\begin{pmatrix} \phi \\ \varphi \end{pmatrix}\,.\label{bosonic}
\ee
In this form, the magnetic scalar Lagrangian density \eqref{magn}, up to total derivatives, becomes
\be
\mathcal{L}_\mathrm{mC}=-\,\frac{\,1\,}{2}\,\mathrm{\Phi}\,\mathrm{B}\,\mathrm{\Phi}\,.\label{Matrix boson}
\ee

\subsection{Lagrange multiplier method}\label{lm}

Another way to obtain Carroll scalar Lagrangians is by using the Lagrange multiplier method, as presented in \cite{deBoer:2021jej}. For this purpose, let us consider two equivalent Lagrangian densities for a massless scalar field, given by \eqref{Scalar Lagrangian} and
\begin{align}
\mathcal{L}'&=-\,\frac{1}{2}\,c^2\,\chi^2+\chi\,\p_t\phi-\,\frac{1}{2}\,(\p_i\phi)^2\,,\label{L'}
\end{align}
where $\chi$ in the latter is a Lagrange multiplier. Indeed, the variation of $\mathcal{L}'$ with respect to $\chi$ equal to zero leads to the equation of motion of $\chi$, which is $c^2\,\chi-\p_t\phi=0$. Solving for $\chi$ and substituting back into $\mathcal{L}'$ one finds $\mathcal{L}'=\mathcal{L}$. Accordingly, two equivalent expressions for the scalar field Lagrangian density allow us to obtain the Carroll limit, $c\rightarrow 0$, in two different ways. First, we may directly take the limit of the visible $c$ in the Lagrangian $\mathcal{L}'$ \eqref{L'}, without rescaling the field. This yields the magnetic Carroll scalar Lagrangian density
\be
\mathcal{L}_\mathrm{mC}= \chi\,\p_t\phi-\,\frac{1}{2}\,(\p_i\phi)^2\,.\label{magnetic Carroll La}
\ee
We note that the Lagrange multiplier $\chi$ in this case can be understood as the momentum of $\varphi$, $\chi=\pi_\varphi$, in the Hamiltonian action \eqref{scalar magnetic}, or equivalently as its time derivative, $\chi=\p_t\varphi$, in the Lagrangian \eqref{magn}.

\vspace{.3cm}

Alternatively, one can rescale the field in the Lagrangian $\mathcal{L}$ \eqref{Scalar Lagrangian}, i.e. $\phi \rightarrow c\,\phi$, and then take the limit $c\rightarrow 0$. This leads to the electric Carroll scalar Lagrangian density
\be
\mathcal{L}_\mathrm{eC}= \frac{1}{2}\,(\p_t\phi)^2\,. \label{electric Carroll La}
\ee

Using the Lagrange multiplier method, we can see that the electric \eqref{electric Carroll La} and magnetic \eqref{magnetic Carroll La} Carroll scalar Lagrangians are invariant (up to total derivatives) under the Carroll boost transformations, which have the following form
\begin{align}
    \d^{^{(\rm{C})}}_{B}\phi={b}^i\,{x}_i\,\p_t\,\phi\,,\qquad\qquad
    \d^{^{(\rm{C})}}_{B}\chi={b}^i\,{x}_i\,\p_t\,\chi\,+\,{b}^i\,{\p}_i\,\phi\,. \label{Carroll trans11}
\end{align}

We observe that the magnetic Carroll Lagrangian \eqref{magnetic Carroll La} preserves its Carroll boost invariance when we set $\p_i\,\phi=0$. Such a Lagrangian, called the ``simplest'' magnetic Lagrangian in \cite{deBoer:2023fnj}, would be invariant under \eqref{Carroll trans11} by setting $\p_i\,\phi=0$. We have generalized this case and its fermionic counterpart to a supersymmetric theory with C-supersymmetry, as shown in appendix \ref{appD}.

\subsection{Seed Lagrange method}\label{sl}

The seed Lagrangian method is an alternative approach to obtain the Carroll scalar field theories \cite{Bergshoeff:2022qkx}. Starting from the Lagrangian density of a massless scalar field \eqref{Scalar Lagrangian}, we can rescale the field as $\phi\rightarrow c\,\phi$, and take the Carroll limit $c\rightarrow 0$. This led to the electric Carroll scalar Lagrangian \eqref{electric Carroll La} which is invariant up to total derivatives under the Carroll boost transformation
\be
\d_B^{^{{(\rm{C})}}}\phi={b}^i\,{x}_i\,\p_t\,\phi\,.\label{cbt}
\ee
To find the magnetic Carroll scalar Lagrangian, one can use the magnetic Galilei scalar Lagrangian (discussed in the next section)
\be
\mathcal{L}_\mathrm{mG}=-\,\frac{1}{2}\,(\p_a\phi)^2\,,\label{seed}
\ee
as the seed Lagrangian. Obviously, the seed Lagrangian \eqref{seed} does not preserve the symmetry under the Carroll boost transformation \eqref{cbt} and it transforms as
\be
\d_B^{^{{(\rm{C})}}}\mathcal{L}_\mathrm{mG}=\phi \,\,b^i\,\p_i\,\p_t\,\phi\,.
\ee
However, the seed Lagrangian can be compensated by the addition of a second term in the Lagrangian which involves a Lagrange multiplier $\chi$
\be
\mathcal{L}^{\chi}_\mathrm{mG}=\chi\,\p_t\,\phi\,.
\ee
This implies that the magnetic Carroll scalar Lagrangian can be obtained as
\be
\mathcal{L}_\mathrm{mC}\,=\,\mathcal{L}_\mathrm{mG}\,+\,\mathcal{L}^\chi_\mathrm{mG}\,=\,-\,\frac{1}{2}\,(\p_a\phi)^2+\,\chi\,\p_t\,\phi\,.\label{mcL}
\ee
We then impose the symmetry of the Lagrangian \eqref{mcL} under the Carroll boost transformation \eqref{cbt}. This allows us to find the transformation of $\chi$ as
\be
\d_B^{^{{(\rm{C})}}}\chi=b^i\,x_i\,\p_t\,\chi+b^i\,\p_i\,\phi\,.\label{cbtx}
\ee
Hence, the Lagrangian \eqref{mcL} is invariant under the Carroll boost transformations \eqref{cbt} and \eqref{cbtx} up to total derivatives.

\section{Galilei scalar field} \label{Gscalar}

The Galilei scalar field theories (electric and magnetic) have been formulated solely through the seed Lagrangian method \cite{Bergshoeff:2022qkx}. In this section, we not only provide a review of the seed Lagrangian method in \ref{eGs}, but also motivated by \cite{Henneaux:2021yzg, deBoer:2021jej}, present three additional approaches for deriving the Galilei scalar Lagrangians: the Hamiltonian formulation method, the field expansion approach, and the Lagrange multiplier method. These methods have different procedures, but they all lead to the equivalent results.

\subsection{Hamiltonian method} \label{HmG}

The Hamiltonian method that applies to Carroll scalars theories \cite{Henneaux:2021yzg} is ineffective for the construction of Galilei scalar theories. Indeed, when we take the Galilei limit, $c\rightarrow\infty$, the field $\pi_\phi$ becomes zero according to \eqref{conjugate b}. This leads to an indeterminate form of $(0\times\infty)$ in the Hamiltonian \eqref{Hamiltonian b} and in the middle term of the action \eqref{scalar Hami}. Therefore, to avoid this ambiguity, we establish an alternative formulation, called ``Hamiltonian-like'' formulation, in which the Galilei scalar field action is well-defined in the limit $c\rightarrow\infty$. To this end, considering the Lagrangian density of a massless scalar field \eqref{Scalar Lagrangian}, we just define a new canonical momentum conjugate to the field $\phi$, subject to the spatial derivatives of $\phi$, which is
\be
\Pi^i_\phi:=\frac{\p\mathcal{L}}{\p\mathring{\phi}_i}=-\,\mathring{\phi}_i\,, \qquad\quad \mathring{\phi}_i\equiv\p_i\phi\,.
\label{conjugate b2}
\ee
Then, the Hamiltonian-like density, which is different from the Hamiltonian density \eqref{Hamiltonian b}, defines
\be
\mathscr{H}=\Pi^i_\phi\,\mathring{\phi}_i-\mathcal{L}=-\,\frac{1}{2}\,\Big[\,(\Pi^i_\phi)^2+\,\frac{1}{c^2}\,(\p_t\phi)^2 \,\Big]\,,
\ee
and the action of a massless scalar field in the Hamiltonian-like formulation becomes
\be
S\,[\phi,\Pi^i_\phi]=\int\,dt\,d^3x\,\Big\{\,\Pi^i_\phi\,\mathring{\phi}_i\,-\, \mathscr{H} \,\Big\}=\int\,dt\,d^3x\,\Big\{ \,\Pi^i_\phi\,\mathring{\phi}_i\,+\,\frac{1}{2}\,(\Pi^i_\phi)^2\,+\,\frac{1}{2c^2}\,(\p_t\phi)^2\,\Big\}\,.
\label{scalar Hami1}
\ee
We observe that the limit $c\rightarrow\infty$ becomes well-defined in this new formulation. Accordingly, we can first take the limit without rescaling the fields, which also does not cause any singularity in \eqref{conjugate b2}. This leads to the action of magnetic Galilei scalar field in the Hamiltonian-like formulation
\be
S_\mathrm{mG}=\int\,dt\,d^3x\,\Big\{ \,\Pi^i_\phi\,\mathring{\phi}_i\,+\,\frac{1}{2}\,(\Pi^i_\phi)^2\,\Big\}\,.
\label{scalar magnetic 1}
\ee
Alternatively, we can rescale the fields in \eqref{scalar Hami1} as follows: $\phi\rightarrow c\, \phi$ and $\Pi^i_\phi\rightarrow \tfrac{1}{c}\,\Pi^i_\phi$. Then, we apply the limit and replace $\Pi^i_\phi$ with the auxiliary field $\Pi^i_\varphi$, since the rescaling makes the canonical relation \eqref{conjugate b2} singular in the limit. This arrives us at the action of electric Galilei scalar field in the Hamiltonian-like formulation
\be
S_\mathrm{eG}=\int\,dt\,d^3x\,\le\{ \,\Pi^i_\varphi\,\mathring{\phi}_i\,+\,\frac{1}{2}\,(\p_t\phi)^2\,\ri\}\,.\label{scalar electric 1}
\ee

\subsection{Field expansion method} \label{fG1}

The derivation of the Galilei scalar theory can be achieved by utilizing the field expansion method analogously to the Carroll scalar theory \cite{deBoer:2021jej}. To demonstrate this, we consider the following Lagrangian density of the massless scalar field, which involves the speed of light $c$\,:
\begin{align}
\mathscr{L}&=-\,\frac{1}{2}\,\p^\m{\varPhi}\,\p_\m{\varPhi}=\frac{1}{2c^2}\,(\p_t{\varPhi})^2-\frac{1}{2}\,(\p_a{\varPhi})^2\,.\label{scalar2}
\end{align}
If we then make an expansion around $c=\infty$ in the scalar field\footnote{Similar to the Carroll scalar case \eqref{Phi expansion}, Galilei boost invariant terms are not influenced by odd powers of $c$.}
\be
{\varPhi}=c^{-\a}\Big(\phi+c^{-2}\,\varphi+\mathcal{O}(c^{-4})\Big)\,,\label{Phi expansion2}
\ee
for some $\a$, and define $\mathcal{L}_\mathrm{mG}$ and $\mathcal{L}_\mathrm{eG}$ through the Lagrangian density
\be
\mathscr{L}=c^{-2\a}\Big( \mathcal{L}_\mathrm{mG}+c^{-2}\,\mathcal{L}_\mathrm{eG}+\mathcal{O}(c^{-4})\Big)\,,
\ee
one arrives respectively at the magnetic and electric Galilei scalar field Lagrangian densities
\begin{align}
     \mathcal{L}_\mathrm{mG}&=-\,\frac{1}{2}\,(\p_a\phi)^2\,,\label{elec2}\\[8pt]
      \mathcal{L}_\mathrm{eG}&=\frac{1}{2}(\p_t\phi)^2-(\p^a\varphi)(\p_a\phi)\,.\label{magn2}
\end{align}

Let us next provide the invariance of these Lagrangians under the Galilei boost. For this purpose, we apply the scalar field expansion \eqref{Phi expansion2} for \eqref{Lorentz tr} and introduce $\vec{\b}=\frac{1}{c}\,\vec{b}$, with $\vec{b}$ being the Galilei boost parameter. This yields the Galilei boost transformations
\begin{align}
    \d_B^{^{(\rm{G})}}\phi=t\,{b}^i\,{\p}_i\,\phi\,,\qquad\qquad
    \d_B^{^{(\rm{G})}}\varphi=t\,{b}^i\,{\p}_i\,\varphi+{b}^i\,{x}_i\,\p_t\,\phi\,, \label{Galilei trans}
\end{align}
under which the magnetic \eqref{elec2} and the electric \eqref{magn2} Galilei scalar Lagrangians are invariant up to total derivatives.

\vspace{.3cm}

The magnetic Galilei Lagrangian \eqref{elec2} yields the equation of motion $\p_i^2\phi=0$. For the electric Galilei Lagrangian \eqref{magn2}, the equations of motion are $\p_i^2\phi=0$ and $\p_i^2\varphi-\p_t^2\phi=0$. These can be compactly written as
\be
 \qquad \mathrm{B}\,\mathrm{\Phi}=0\,, \quad\qquad \mbox{with}\quad \mathrm{B}:=\begin{pmatrix} -\,\p_t^2 & \p_i^2 \\ ~~\,\p_i^2 & 0  \end{pmatrix}\,,\quad \mathrm{\Phi}:=\begin{pmatrix} \phi \\ \varphi \end{pmatrix}\,.\label{bosonic g}
\ee
The electric Galilei scalar Lagrangian \eqref{magn2}, up to total derivatives, can then be expressed as
\be
\mathcal{L}_\mathrm{eG}=\frac{\,1\,}{2}\,\mathrm{\Phi}\,\mathrm{B}\,\mathrm{\Phi}\,.\label{Matrix boson g}
\ee

\subsection{Lagrange multiplier method}

The Lagrange multiplier method for Carroll scalar theory \cite{deBoer:2021jej} included a Lagrangian \eqref{L'} that was well-defined in the Carroll limit. However, this Lagrangian becomes singular in the Galilei limit $c\to\infty$. To avoid this, we present and start from an alternative Lagrangian which is well-defined in the Galilei limit. To this end, we employ a Lagrange multiplier $\chi^a$ and consider the following Lagrangian density for a massless real scalar field:
\begin{align}
\mathcal{L}'&=\frac{1}{2c^2}\,\chi^a\chi_a - \chi^a\,(\p_a\phi)+\frac{1}{2}\,(\p_t\phi)^2\,.\label{L' Gal}
\end{align}
This Lagrangian density is equivalent to the standard one
\be
\mathcal{L}=\frac{1}{2c^2}\,(\p_t\phi)^2-\frac{1}{2}\,(\p_a\phi)^2\,, \label{Scalar Lagrangian2}
\ee
after rescaling the field as $\phi \rightarrow \frac{1}{c}\,\phi$. This can be seen by eliminating the Lagrange multiplier $\chi^a$ using its equation of motion, which gives $\chi_a = c^2 \,\p_a \phi$. Substituting this back into the Lagrangian density \eqref{L' Gal}, and rescaling the field $\phi \rightarrow \frac{1}{c}\,\phi$, we obtain \eqref{Scalar Lagrangian2} which is the usual form of the Lagrangian density for a massless scalar field. Having these two Lagrangians, one can take the Galilean limit, $c\rightarrow \infty$, explicitly. If we do this, the Lagrangian \eqref{L' Gal} reduces to the electric Galilei scalar Lagrangian
\be
\mathcal{L}_\mathrm{eG}=\frac{1}{2}\,(\p_t\phi)^2- \chi^a\,(\p_a\phi)\,,\label{electric Galil}
\ee
while \eqref{Scalar Lagrangian2} yields the magnetic Galilei scalar Lagrangian
\be
\mathcal{L}_\mathrm{mG}=-\,\frac{1}{2}\,(\p_a\phi)^2\,. \label{magnetic Galil}
\ee
These electric and magnetic Lagrangians are identical to those obtained in \cite{Bergshoeff:2022qkx} by using the seed Lagrangian method. Furthermore, we observe that the Lagrange multiplier $\chi^a$ in this case has the interpretation of either the negative of the momentum of $\varphi$, $\chi^a=-\,\Pi^a_\varphi$, in the Hamiltonian action \eqref{scalar electric 1}, or the spatial derivative of $\varphi$, $\chi^a=\p^a\varphi$, in the Lagrangian \eqref{magn2}.

\subsection{Seed Lagrangian method}\label{eGs}

We briefly review the seed Lagrangian method, presented in \cite{Bergshoeff:2022qkx}, as another technique to derive the Galilei scalar field Lagrangians. We start from the relativistic massless scalar field theory with the Lagrangian density \eqref{Scalar Lagrangian2} and take the limit $c\rightarrow\infty$, which corresponds to the Galilei limit. This gives us the magnetic Galilei scalar Lagrangian density \eqref{magnetic Galil}, which is invariant under the Galilean boost transformation
\be
\d_B^{^{{(\rm{G})}}}\phi=t\,b^i\,\p_i\,\phi\,.\label{gbt}
\ee
On the other side, to derive the electric Galilei scalar Lagrangian density, one can utilize the electric Carroll scalar Lagrangian density \eqref{elec} as a seed Lagrangian, that is
\be
\mathcal{L}_\mathrm{eC}=\frac{1}{2}\,(\p_t\phi)^2\,.\label{seed g}
\ee
This choice of the seed Lagrangian is not invariant under the Galilean boost transformation \eqref{gbt}. It transforms as
\be
\d_B^{^{{(\rm{G})}}}\mathcal{L}_\mathrm{eC}=-\,\phi\,\p_t\,b^i\,\p_i\,\phi\,.
\ee
In order to obtain a Galilei boost invariant Lagrangian density, we can compensate the seed Lagrangian \eqref{seed g} by the addition of a second term in the Lagrangian which involves a Lagrange multiplier $\chi^a$
\be
\mathcal{L}^\chi_\mathrm{eC}=-\,\chi^a\,\p_a\,\phi\,.
\ee
As a consequence, the electric Galilei scalar Lagrangian density can be found as
\be
\mathcal{L}_\mathrm{eG}\,=\,\mathcal{L}_\mathrm{eC}\,+\,\mathcal{L}^\chi_\mathrm{eC}\,=\,
\frac{1}{2}\,(\p_t\phi)^2\,-\,\chi^a\,\p_a\,\phi\,. \label{eGL}
\ee
Imposing the symmetry of this Lagrangian under the Galilei boost transformation \eqref{gbt}, one can determine the transformation of $\chi^a$, which becomes
\be
\d_B^{^{{(\rm{G})}}}\chi^a=t\,b^i\,\p_i\,\chi^a\,+\,b^a\,\p_t\,\phi\,.\label{gbt2}
\ee
Therefore, the obtained electric Galilei scalar Lagrangian density \eqref{eGL} is invariant under the Galilean boost transformations \eqref{gbt} and \eqref{gbt2}, up to total derivatives.

\section{Simplest Carroll supersymmetric theory} \label{appD}

The theory of the electric Carroll scalar field is characterized by a straightforward Lagrangian, as represented by relation \eqref{electric Carroll La}. However, the Lagrangian density for the magnetic Carroll scalar field has more terms, involving spatial derivative terms and a Lagrange multiplier term \eqref{magnetic Carroll La}. Nonetheless, it is possible to consider a simpler Lagrangian density for the magnetic Carroll scalar field. In this simplified Lagrangian, the spatial derivatives of the field are eliminated, leaving only the Lagrange multiplier term intact. This Lagrangian, which still exhibits Carroll boost invariance, has been explored in \cite{deBoer:2023fnj} and called the ``simplest'' magnetic Carroll Lagrangian, given by
\be
\mathcal{L}^{^\mathrm{\,S}}_\mathrm{mC}= \chi\,\p_t\,\phi\,.\label{simp1}
\ee
Similarly, by eliminating the spatial derivatives of the field in \eqref{magn f}, and defining $\lambda:=\p_t\,\eta$ for consistency of the supersymmetric theory, we can present the simplest Carroll boost invariant magnetic Lagrangian for fermions, which is given by
\be
\mathcal{L}^{^\mathrm{\,S}}_\mathrm{mC}=\bar{\lambda}\,\g^0\,\psi\,-\,\bar{\psi}\,\g^0\,\lambda\,.\label{simp2}
\ee

In this section, we apply the simplest Lagrangians for Carroll fields to construct the simplest supersymmetric theory for the magnetic Carroll case with $\mathcal{N}=1$ off-shell supersymmetry. We find such an action can be given by
\begin{align}
S^{^{^\mathrm{\,S-SUSY}}}_\mathrm{mC}&=\int dt\,d^3x\,
\Big\{\,\chi_1\,\p_t\,\phi_R\,+\,\chi_2\,\p_t\,\phi_I\,+\,F_R\,G_R\,+\,F_I\,G_I\,+\,\frac{1}{2}\,\bar\lambda\,\g^0\,\psi\,-\,\frac{1}{2}\,\bar\psi\,\g^0\,\lambda \,\Big\}\,.
\label{Simplestsuper-magn action}
\end{align}
The bosonic part of the action contains two real scalar fields, $\phi_R$ and $\phi_I$, which are the real and imaginary parts of $\phi$, and two Lagrange multipliers $\chi_1$ and $\chi_2$. It also contains four real auxiliary scalar fields: $F_R$ and $F_I$, which are the real and imaginary parts of $F$, and $G_R$ and $G_I$, which are the real and imaginary parts of $G$. These auxiliary fields have no dynamics, but they are necessary for the off-shell supersymmetry. The fermionic part of the action includes a Majorana spinor field $\psi$ and the fermionic Lagrange multiplier $\lambda$. We find that the action \eqref{Simplestsuper-magn action} is invariant under the following magnetic Carroll supersymmetry transformations
\begin{subequations}
\label{super-magn tr1}
\begin{align}
     \d\phi_R&=\bar\ep\,\psi \,, ~\,\hspace{7cm}\d\phi_I=\bar\ep\,i\,\g^5\,\psi \,,\\[8pt]
     \d\chi_1&=\bar\ep\,\lambda \,, ~\,\,\hspace{7cm}\d\chi_2=\bar\ep\,i\,\g^5\,\lambda \,,\\[8pt]
     \d F_R&=-\,\bar\ep\,\g^0\,\p_t\,\psi \,, \hspace{6cm}\d F_I=\bar\ep\,i\,\g^0\g^5\,\p_t\,\psi  \,,\\[8pt]
     \d G_R&=-\,\bar\ep\,\g^0\,\lambda \,, \,~\,\hspace{6cm}\,\d G_I=\bar\ep\,i\,\g^0\g^5\,\lambda  \,,\\[8pt]
    \d\psi&=\g^0\p_t\,(\,\phi_R+i\,\g^5\,\phi_I\,)\,\ep- (\,F_R+i\,\g^5\,F_I\,)\,\ep\,,\\[8pt]
    \d\lambda&=\g^0\p_t\,(\,\chi_1+i\,\g^5\,\chi_2\,)\,\ep- \,\p_t\,(\,G_R+i\,\g^5\,G_I\,)\,\ep\,.
\end{align}
\end{subequations}
\noindent We can then verify that the commutator of these transformations on each field yields
\be
[\,\d\1\,,\,\d\2\,]=2\,(\bar\ep\2\,\g^0\,\ep\1)\,\p_t\,,\label{csusy}
\ee
which demonstrates that the supersymmetry algebra closes off-shell. It is important to remark that the super magnetic Carroll theory does not possess C-supersymmetry, but rather G-supersymmetry as specified by \eqref{gsusy}. However, when we consider the simplest Lagrangians for the Carroll fields, given by \eqref{simp1} and \eqref{simp2}, we can have a C-supersymmetry \eqref{csusy} rather than a G-supersymmetry.

\vspace{.3cm}

We could also introduce the simplest electric Galilei Lagrangians and build their supersymmetric extension, but for now, we will refrain from doing so, as the process follows a similarly straightforward method.

{\small
\bibliographystyle{hephys}
\bibliography{references}
}


\end{document}